\documentclass[preprint,12pt]{elsarticle}

\usepackage{amssymb}
\usepackage{url} 
\usepackage{amsmath}
\usepackage{multirow}
\usepackage{threeparttable}
\usepackage{color}

\journal{}

\begin{document}

\begin{frontmatter}



\title{UniASM: Binary Code Similarity Detection without Fine-tuning}


\author[inst1]{Yeming Gu}
\author[inst1]{Hui Shu}
\author[inst1]{Fei Kang}
\author[inst1]{Fan Hu}

\affiliation[inst1]{organization={Key Laboratory of Cyberspace Security, Ministry of Education},%Department and Organization
            city={Zhengzhou},
            country={China}}

\begin{abstract}
Binary code similarity detection (BCSD) is widely used in various binary analysis tasks such as vulnerability search, malware detection, clone detection, and patch analysis. Recent studies have shown that the learning-based binary code embedding models perform better than the traditional feature-based approaches. However, previous studies have not delved deeply into the key factors that affect model performance. In this paper, we design extensive ablation studies to explore these influencing factors. The experimental results have provided us with many new insights. We have made innovations in both code representation and model selection: we propose a novel rich-semantic function representation technique to ensure the model captures the intricate nuances of binary code, and we introduce the first UniLM-based binary code embedding model, named UniASM, which includes two newly designed training tasks to learn representations of binary functions. The experimental results show that UniASM outperforms the state-of-the-art (SOTA) approaches on the evaluation datasets. The average scores of Recall@1 on cross-compilers, cross-optimization-levels, and cross-obfuscations have improved by 12.7\%, 8.5\%, and 22.3\%, respectively, compared to the best of the baseline methods. Besides, in the real-world task of known vulnerability search, UniASM outperforms all the current baselines.
\end{abstract}

\begin{graphicalabstract}
\includegraphics[width=5.4in]{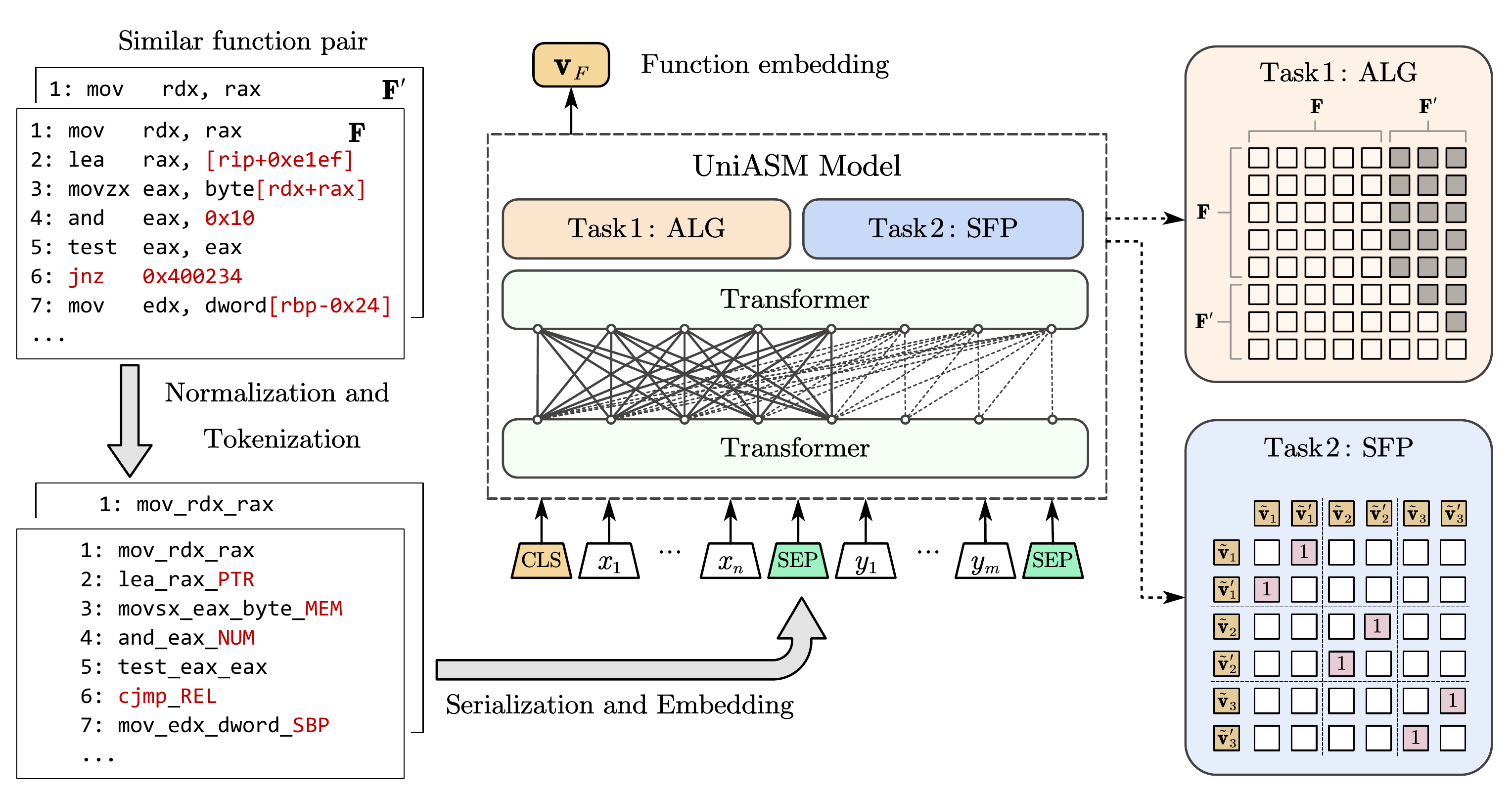}
\end{graphicalabstract}

\begin{highlights}
\item We propose a novel assembly language processing model, UniASM, the first UniLM-based model for BCSD. Our model outperforms the baselines and can be used in the real-world vulnerability search task. We have released the code and the pre-trained model of UniASM at \url{https://github.com/clm07/UniASM}. 
\item We propose a novel rich-semantic function representation technique, which retains a wealth of semantic information, ensuring that the model captures the intricate nuances of binary code.
\item We design an extensive suite of ablation studies to delve deeply into the various factors influencing the model’s accuracy in BCSD tasks, yielding many inspiring findings. 
\end{highlights}


\end{frontmatter}


\section{Introduction}\label{sec:introduction}

Binary code similarity detection (BCSD) is widely used in vulnerability search \cite{Liu2018alphaDC, Zuo2019NeuralMT}, malware detection \cite{Cesare2011MalwareVD, Cesare2014ControlFM, Tams2021SIMBIoTASM}, clone detection \cite{Hu2017BinaryCC, Ding2016Kam1n0MA}, patch analysis \cite{Xu2017SPAINSP}, etc. Most commercial software is closed-sourced and consists of a large amount of binary code. Therefore, the study of BCSD has crucial practical significance.
One of the main challenges of BCSD is that different compilers, optimization levels, or code obfuscations can lead to significant changes in the binary code. The target binaries lose most of the natural semantic information of the source code during the compilation process. Since the binary code does not have vocabularies containing natural semantics as the source code, extracting semantic features from it is challenging. Sæbjørns et al. \cite{Sbjrnsen2009DetectingCC} try to extract statistical features of instructions for BCSD manually. However, the statistical characteristics vary with the compilation optimization options, resulting in a degradation of BCSD performance. Other works \cite{Ji2021BugGraphDS, Xu2017NeuralNG} try to analyze similarity through the control flow graph (CFG). However, different compile options or code obfuscations may lead to different CFGs.

As none of the traditional similarity comparison methods have addressed the problem of cross-optimization levels and cross-obfuscations, the deep learning-based models are considered promising candidate methods for BCSD. In recent years, natural language processing (NLP) models have shown their capabilities of semantic understanding and text embedding. The state-of-the-art research in BCSD has begun to employ NLP models. Asm2vec \cite{Ding2019Asm2VecBS} generates the embedding of instructions and functions based on the PV-DM model \cite{Le2014DistributedRO}. SAFE \cite{Massarelli2019SAFESF} uses the skip-gram method \cite{TensorFlow2022Word2Vec} and self-attention network \cite{Lin2017ASS} to generate the embedding. However, neither PV-DM nor skip-gram can learn the complex semantic features of the binary code because they rely heavily on instructions of similarity in the binary code pairs. Recent studies try to use more complex models: PalmTree \cite{Li2021PalmTreeLA} is the first to apply BERT \cite{Devlin2019BERTPO} to instruction embedding, and jTrans \cite{Wang2022jTransJT} leverages BERT to learn the control flow information of the functions. They achieved better performance than traditional methods. However, some key issues need to be studied in depth:

\begin{itemize}
    \item Which backbone model should be chosen for binary code embedding? 
    \item What training tasks are better for BCSD? 
    \item How to serialize the assembly code properly? 
\end{itemize}

This paper proposes a toolkit called UniASM, designed to achieve high BCSD performance and can be used directly without fine-tuning. We have designed two training tasks for UniASM: Assembly Language Generation (ALG, Section \ref{sec:ALG}) and Similar Function Prediction (SFP, Section \ref{sec:SFP}). ALG predicts the second function in the input sequence based on unidirectional attention, while SFP predicts the similarity of the two functions in the input sequence. After training, the generated function embeddings can be used for BCSD tasks directly.

The contributions of this paper are summarized as follows:

\begin{itemize}
    \item We propose a novel assembly language processing model, UniASM, the first UniLM-based \cite{Dong2019UnifiedLM} model for BCSD. Our model outperforms the baselines and can be used in the real-world vulnerability search task. We have released the code and the pre-trained model of UniASM at \url{https://github.com/clm07/UniASM}. 
    
    \item We propose a novel rich-semantic function representation technique (Section \ref{sec:norm token}), which retains a wealth of semantic information, ensuring that the model captures the intricate nuances of binary code.
    
    \item We design an extensive suite of ablation studies to delve deeply into the various factors influencing the model’s accuracy in BCSD tasks, yielding many inspiring findings:

    \begin{enumerate}
    \item The UniLM-based model achieves high performance in BCSD tasks and is significantly better than the BERT-based model (Section \ref{sec:abla models}). 
    \item The pre-training task ALG is more suitable for BCSD than the widely used MLM (Section \ref{sec:abla tasks}). 
    \item Full-instruction tokenization shows better performance than fine-grained algorithms (Section \ref{sec:abla token}).
    \item Neither random-walk nor longest-walk performs any better than the linear serialization of a function (Section \ref{sec:abla corpora}).    
    \item Transformer-based function similarity analysis does not require a very long input sequence length. A fixed length of 256 is sufficient for achieving good performance (Section \ref{sec:abla seqlen}).
    \end{enumerate}

\end{itemize}

\section{Related Works}
\label{sec:relatedwork}
\subsection{Traditional BCSD Approaches}
\label{sec:traditional appr}

BCSD is one of the popular research areas of binary analysis. Earlier studies tend to implement vectorization of binary codes by extracting dynamic or static features. 

\textbf{Dynamic approaches.} Dynamic methods collect run-time information by executing the program in reality or simulation. BinHunt \cite{GaoRS08BinHunt} and iBinHunt \cite{MingPG12iBinHunt} extract the semantics of functions through symbolic execution and deep taint analysis. However, symbolic execution incurs high costs and is difficult to run on large-scale binaries. The basic idea of Blex \cite{EgeleWCB14Blex}, BinGo \cite{Chandramohan2016BinGoCC}, BinGo-E \cite{XueXCL19BinGoE}, and Multi-MH \cite{Pewny2015CrossarchitectureBS} is to obtain the I/O values of functions by executing the target program. The main shortcoming of these dynamic methods is that the I/O values cannot fully represent the semantics of the function. CACompare \cite{Hu2017BinaryCC} and BinMatch \cite{HuWZLG21BinMatch} leverage emulate executions to obtain richer function semantics to improve similarity comparison performance. IMF-sim \cite{Wang2017InmemoryFF} and BinSim \cite{MingXJW17BinSim} use finer-grained run-time features to identify differences between two execution traces. Dynamic approaches can obtain additional run-time features, such as parameters, I/O values, and execution traces. However, they are computationally expensive and require a complex analysis environment, which limits practical usage.

\textbf{Static approaches.} Static features such as instructions, basic blocks, function calls, and control flow are used to achieve similarity comparison. IDA FLIRT \cite{Hexrays2022FLIRT} and UNSTRIP \cite{Jacobson2011LabelingLF} identify library functions by generating fingerprints statically. BinClone \cite{FarhadiFCD14BinClone}, ILINE \cite{JangWB13ILINE}, MutantX-S \cite{HuSBG13MutantX}, BinSign \cite{NouhRMDH17BinSign}, and BinShape \cite{Shirani0D17BinShape} use statistical features of binary code to achieve similarity analysis. Tracelet \cite{DavidY14Tracelet} and BinSequence \cite{HuangYD17BinSeq} focus on instruction sequences and use edit distance to compare two instruction sequences. ESH combines the similarity of code fragments to ultimately measure the similarity between procedures. In order to better utilize the control flow information of functions, TEDEM \cite{PewnySBHR14TEDEM}, XMATCH \cite{FengWZZHY17XMATCH}, and Sæbjørns et al. \cite{Sbjrnsen2009DetectingCC} use tree edit distance or graph edit distance to compare the CFGs of functions. However, comparison based on edit distance is computationally complex and sensitive to structural changes. To address this, DiscovRe \cite{Eschweiler2016discovREEC}, BinDiff \cite{Dullien2005GraphbasedCO}, Genius \cite{Feng2016ScalableGB} and Kam1n0 \cite{Ding2016Kam1n0MA} leverage graph isomorphism instead of comparing edit distance to improve the efficiency of graph comparison. The disadvantage of graph isomorphism is that it requires high-quality node features.

\subsection{Learning-based BCSD Approaches}
\label{sec:learning appr}

Deep learning has achieved satisfactory results in tasks such as image processing and language understanding. One of the popular research directions in deep learning, embedding, transforms inputs into low-dimensional dense vectors, which can be conveniently applied to various downstream tasks. Early works, such as word2vec \cite{Mikolov2013DistributedRO} and GloVe \cite{PenningtonSM14GloVe}, can generate vectors for word representation. BERT \cite{Devlin2019BERTPO} uses pre-training and fine-tuning to accomplish downstream tasks such as text classification. However, its generated sentence embedding performs poorly when directly used for those tasks \cite{ReimersG19SBERT}. To address this, some studies, such as SimCSE \cite{GaoYC21SimCSE} and Mirror-BERT \cite{VKC21MirrorBERT}, leverage Siamese networks to improve embedding performance for downstream tasks. Other works, such as DPR \cite{KarpukhinOMLWEC20DPR}, Condenser \cite{GaoC21Condenser}, and GPL \cite{WangT0G22GPL}, aim to generate better embeddings for dense retrieval. In addition, generative models such as GPT \cite{OpenAI2018GPT} and UniLM \cite{Dong2019UnifiedLM} also show great potential in text embedding. In binary analysis research, recent studies have started to apply learning-based methods to BCSD tasks.

\textbf{DNN-based approaches.} Deep Neural Network (DNN) is a multi-layer neural network mainly used to process images, audio, and text. Inspired by image processing, Marastoni et al. \cite{MarastoniGP18} translated binaries into images and used Convolutional Neural Networks (CNN) to process the generated images, achieving program classification. However, this method can only be applied to small binaries because the CNN network needs to see the entire binary image. $\alpha$diff \cite{Liu2018alphaDC} works on the function instead of the whole binary and learns function embeddings directly from the sequence of raw bytes using CNN. VulSeeker \cite{GaoYFJS18VulSeeker} extracts basic block features and inputs them into a DNN to generate function embeddings for vulnerability function search. DNN-based methods cannot handle the order information of input data well, while the execution order is crucial to the code semantics.

\textbf{Graph-based approaches.} Graph Neural Network (GNN) can directly process graph data and can be used to learn program semantics from control flow, data flow, and function call relationships. Gemini \cite{Xu2017NeuralNG} and GraphEmb \cite{Massarelli2019InvestigatingGE} extract attributed control flow graphs (ACFGs) for functions and train a graph embedding network to generate embeddings. GMNN \cite{LiGDVK19GMNN} proposes graph matching networks to generate similarity scores instead of generating embeddings separately to compute the similarity between graphs more efficiently. BugGraph \cite{Ji2021BugGraphDS} utilizes a graph triplet-loss network on the ACFG to produce a similarity ranking. Bin2vec \cite{DArakelyanAHKG21Bin2vec} attempts to use graph convolutional networks to improve the processing performance of graph embeddings. HBinSim \cite{WangJHLH21HBinSim} believes that different features of functions should have different weights in BCSD, so it uses a hierarchical attention graph embedding network to implement ACFG embedding. Asteria \cite{YangCZLZS21Asteria} extracts the syntax tree of functions instead of CFG and uses a Tree-LSTM network to generate function embeddings. However, both graph embeddings and Tree-LSTM face the problem of high computational complexity for large-scale graph data and heavily rely on the accuracy of node features.

\textbf{NLP-based approaches.} Natural Language Processing (NLP) has shown excellent performance in text processing and semantic understanding tasks. It can also be used for semantic learning from binary code by extracting assembly semantics. Asm2vec \cite{Ding2019Asm2VecBS} generates embeddings for instructions and functions using the word2vec model. In addition to word2vec, InnerEye \cite{Zuo2019NeuralMT} utilizes Long Short-Term Memory (LSTM) to learn basic block embeddings. Zhengping Luo et al.’s research \cite{LuoHZZL21} uses a Siamese network to implement similarity comparison of basic block embeddings generated by LSTM. To learn more semantics, Transformer-based models have become a research hotspot in recent years. PalmTree \cite{Li2021PalmTreeLA}, DeepSemantic \cite{DeepSemantic}, and BinShot \cite{AhnAKP22BinShot} apply the BERT model \cite{Devlin2019BERTPO} to binary code embedding and show the great potential of language models in BCSD. MIRROR \cite{Zhang2020SimilarityMM} and CRABS-former \cite{FengLCWF24} aim to cross-architecture similarity analysis by a transformer-based neural machine translation model. Transformer-based approaches require translating instructions or functions into a sequence, which may lead to the loss of function control flow information. To address this, jTrans \cite{Wang2022jTransJT} is the first study to embed control flow information of binary code into Transformer-based language models.

\textbf{Hybrid approaches.} As single methods always have limitations, some studies attempt to mix multiple models to achieve better BCSD performance. SAFE \cite{Massarelli2019SAFESF} uses word2vec to generate instruction embeddings and then utilizes a self-attentive neural network to generate function embeddings. DeepBinDiff \cite{DuanLWY20} first trains a token embedding model derived from word2vec and then leverages the Text-associated DeepWalk \cite{YangLZSC15TADW} algorithm to learn basic block embeddings from the inter-procedural control-flow graphs. BinDNN \cite{LagemanKWM16BinDNN} utilizes three types of neural network models: CNN, LSTM, and regular fully connected feed-forward neural networks. There are other hybrid approaches, such as BEDetector \cite{YuLSHZ21BEDetector}, which combines NLP model and graph auto-encoder model to generate function embeddings, and Codee \cite{YangFLYZ22Codee}, which combines NLP model and network representation learning model. OrderMatters \cite{Yu2020OrderMS} integrates more models, including word2vec, BERT, MPNN \cite{GilmerSRVD17MPNN}, and CNN. UPPC \cite{Zhang22UPPC} used the Siamese network architecture on a combination of word2vec and DPCNN \cite{JohnsonZ17DPCNN}. Although hybrid methods can achieve complementary advantages, they make model training and usage more difficult, and data processing and computational costs are relatively high. 

Overall, Learning-based BCSD methods have better adaptability and performance than traditional methods. Among them, Transformer-based methods show the best potential performance. However, there are significant differences between assembly language and natural language, one of which is the absence of a question-and-answer relationship. This makes it challenging to construct effective datasets for training existing models, such as SimCSE, DPR, GPL, etc. Existing researches have only tried BERT-based methods, and research on model training and data processing is still insufficient. Further research is needed, including backbone models, training tasks, tokenization methods, etc.

\section{Methodology}
\label{sec:mothodology}

\subsection{Overview}
\label{sec:overview}

\begin{figure*}[!t]
\centering
\includegraphics[width=5.4in]{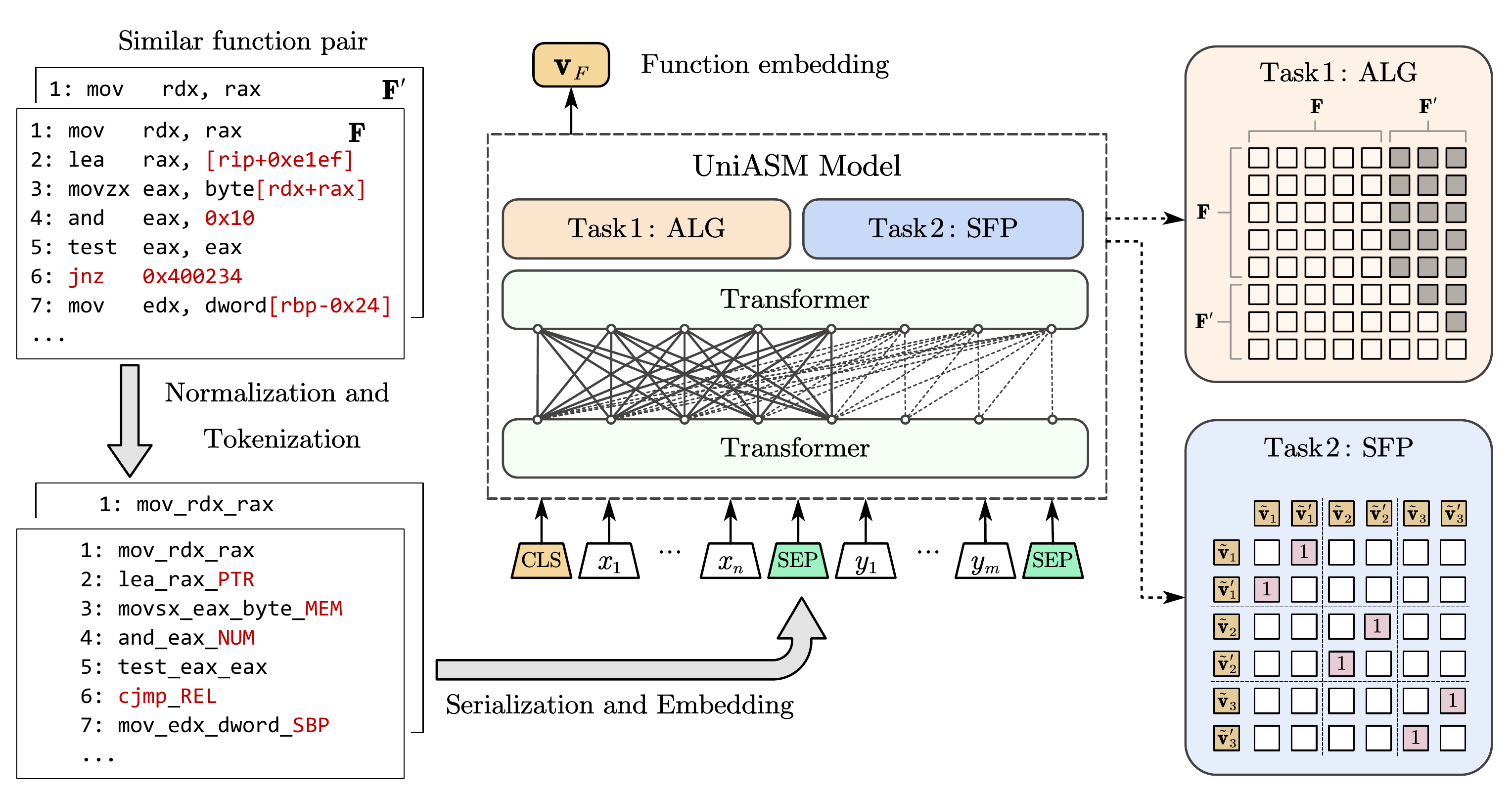}
\caption{Overview of UniASM.}
\label{fig:overview}
\end{figure*}

UniASM is mainly inspired by SimBERT \cite{Su2020SIMBERT} and UniLM \cite{Dong2019UnifiedLM}. UniLM uses bidirectional and unidirectional attention to achieve semantic understanding and generative capabilities. SimBERT proposes a new similarity query task for each batch. UniASM is a transformer-based model and uses two training tasks: Assembly Language Generation (ALG) and Similar Function Prediction (SFP). ALG leverages unidirectional attention to generate the second half of the sequence. SFP is a function query task similar to the query task in SimBERT, which enables the generated function embeddings to be used directly in the BCSD tasks.

Figure \ref{fig:overview} shows an overview of UniASM. For training, the input sequence is constructed from a pair of similar functions. First, the instructions of the functions are normalized to remove the noisy words and mitigate the OOV problem. Then, the instructions are tokenized according to a simple principle: one instruction produces one token. Next, we use a simple linear serialization approach to convert a function into a sequence of tokens. Finally, the sequence is used as the input of UniASM. 

For evaluation, the input sequence is constructed from one function, and the output of the model is the function embedding. We compute the cosine similarity between the two function embeddings as our model-predicted similarity.

\subsection{Function Representation}
\label{sec:norm token}

The raw representation of a binary function is a series of instructions that cannot be used directly. We design a new representation approach for binary functions. It mainly contains three stages: instruction normalization, assembly tokenization, and function serialization.

\subsubsection{Instruction Normalization}
\label{sec:inst rnorm}

Instruction normalization makes instructions look cleaner by replacing the addresses, immediate numbers, float instructions, and conditional jumps. The main principles are as follows:

\begin{itemize}
    \item The indirect addressing with register \textit{eip/rip} is replaced by \textit{PTR}.
    \item The indirect addressing with register \textit{esp/rsp} is replaced by \textit{SSP}.
    \item The indirect addressing with register \textit{ebp/rbp} is replaced by \textit{SBP}.
    \item Other indirect addressing is replaced by \textit{MEM}.
    \item The relevant addressing is replaced by \textit{REL}.
    \item The immediate number is replaced by \textit{NUM}.
    \item The float instruction with register \textit{xmm} is replaced by \textit{XMM}.
    \item The conditional jump, such as \textit{jnz}, is replaced by \textit{cjmp}.
\end{itemize}

The ablation study (Section \ref{sec:abla norm}) shows that this normalization approach can effectively balance token semantics and OOV issues and performs well in the BCSD tasks.

\subsubsection{Assembly Tokenization}
\label{sec:asm token}

Tokenization decomposes unstructured data and texts into chunks of information that can be considered discrete elements called tokens. In this paper, the whole instruction is treated as a token. The advantage is that the instruction contains richer semantic information than individual operands. In practice, we replace the white space with an underline for an instruction, e.g., “\textit{mov rax, 0x10}” will be represented by the token “\textit{mov\_rax\_NUM}.” The ablation study (Section \ref{sec:abla token}) shows that this full-instruction tokenization approach performs much better than the fine-grained approaches.

\subsubsection{Function Serialization}
\label{sec:func serial}

Function serialization aims to serialize the structured function into a sequence of tokens. The approach used in this paper is to serialize the function directly in linear order (address order). Experimental results (Section \ref{sec:abla corpora}) show that linear serialization performs similarly to random-walk and longest-walk. However, random-walk and longest-walk require the construction of a CFG of the function, which is time-consuming. Even worse, longest-walk has to search the longest path on the CFG, which is difficult.

\subsection{Backbone Network}
\label{sec:backbone}

\begin{figure}[!t]
\centering
\includegraphics[width=4in]{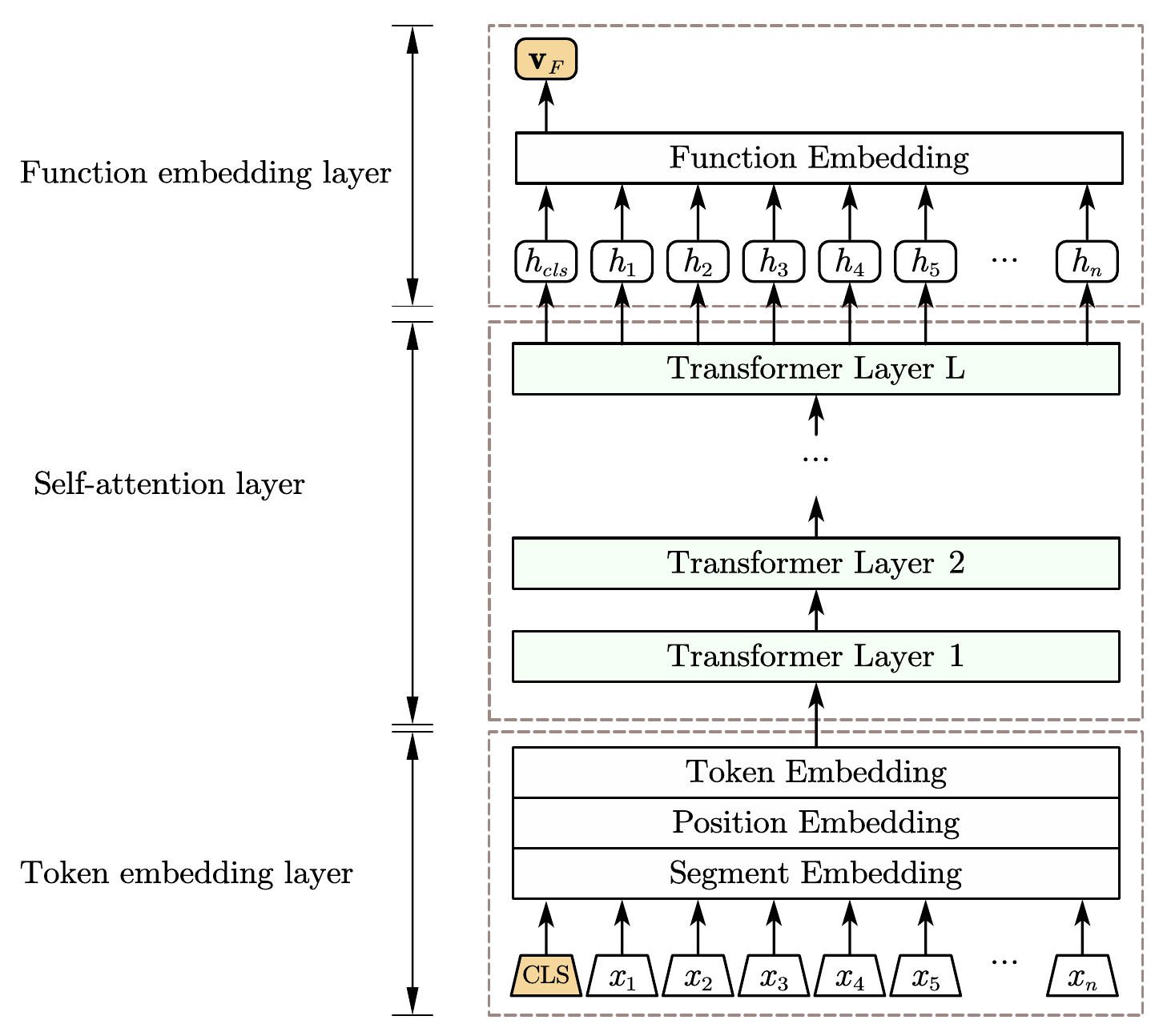}
\caption{Backbone network.}
\label{fig:backbone}
\end{figure}

The base model used in UniASM is a transformer model, as shown in Figure \ref{fig:backbone}, which has shown a strong capability in the representation learning of natural semantics. According to the processing flow, it can be divided into three parts: the token embedding layer, the self-attention layer, and the function embedding layer.

\subsubsection{Token Embedding Layer}
\label{sec:emb layer}

The token embedding layer is used to generate the input vector for the token sequence of the function. For the token sequence of the input function $\mathbf{F}=[ {{x}_{1}},{{x}_{2}},\cdots ,{{x}_{n}} ]$, where ${{x}_{i}}$ represent the $i$-th token of the function, the input vector ${{\mathbf{H}}^{0}}=[ E( {{x}_{1}} ),E( {{x}_{2}} ),\cdots ,E( {{x}_{n}} ) ]$ is obtained by summing the token embedding $E{{x}_{i}}$, position embedding $E{{p}_{i}}$, and segment embedding $E{{s}_{i}}$: 

\begin{equation}
    E( {{x}_{i}} )=E{{x}_{i}}+E{{p}_{i}}+E{{s}_{i}}.
\end{equation}

\subsubsection{Self-attention Layer}
\label{sec:att layer}

The self-attentive layer consists of multiple transformer layers stacked on top of each other, as shown in Figure \ref{fig:backbone}. The input vector ${{\mathbf{H}}^{0}}=[ E( {{x}_{1}} ),E( {{x}_{2}} ),\cdots ,E( {{x}_{n}})]$ is used as the input to the first layer of the Transformer. For the Transformer with the total number of $L$ layers, the output of the $l$-th layer is represented as ${{\mathbf{H}}^{l}}={\textbf{Transformer}_{l}}({{\mathbf{H}}^{l-1}}),l\in [ 1,L ]$, and the self-attention is calculated as follows:

\begin{align}
{{\mathbf{Q}}_{l}} &= {{\mathbf{H}}^{l-1}}\mathbf{W}_{l}^{Q},{{\mathbf{K}}_{l}}={{\mathbf{H}}^{l-1}}\mathbf{W}_{l}^{K},{{\mathbf{V}}_{l}}={{\mathbf{H}}^{l-1}}\mathbf{W}_{l}^{V} \\
%{{\mathbf{M}}_{ij}} &= \left\{ 0,\text{allow to attend} -\infty,\text{prevent from attending}\\
{{\mathbf{M}}_{ij}}&=\left\{ \begin{matrix}
   0, & \text{allow to attend}  \\
   -\infty, & \text{prevent from attending}  \\
\end{matrix} \right.\\
{{\mathbf{A}}_{l}} &= \textbf{softmax}\left( \frac{{{\mathbf{Q}}_{l}}\mathbf{K}_{l}^{\top }}{\sqrt{{{d}_{k}}}}+\mathbf{M} \right){{\mathbf{V}}_{l}} 
\end{align}

The output of the previous layer ${{\mathbf{H}}^{l-1}}$ generates ${{\mathbf{Q}}_{l}}$, ${{\mathbf{K}}_{l}}$ and ${{\mathbf{V}}_{l}}$ through three parameter matrices $\mathbf{W}_{l}^{Q}$, $\mathbf{W}_{l}^{K}$, $\mathbf{W}_{l}^{V}$. The mask matrix ${{\mathbf{M}}_{ij}}$ defines the attention between the tokens. The output ${{\mathbf{A}}_{l}}$ is summed with the ${{\mathbf{H}}^{l-1}}$ residual operation and the feed-forward network finally generates a new hidden layer vector ${{\mathbf{H}}^{l}}$.

\subsubsection{Function Embedding Layer}
\label{sec:func layer}

The function embedding layer generates the embedding vector of the input function. In this paper, we calculate the function embedding vector by the output vector of the token “CLS”:

\begin{equation}
{{\mathbf{v}}_{F}}=\textbf{tanh}( {{h}_{{CLS}}} )\cdot {{\mathbf{W}}^{F}},
\end{equation}

\noindent where $\textbf{tanh}(\cdot)$ is the activation function, ${{\mathbf{W}}^{F}}$ is the parameter matrix of the fully connected network.

\subsection{Training Tasks}
\label{sec:tasks}

UniASM abandons the commonly used mask language model (MLM) and next sentence prediction (NSP) pre-training tasks of BERT in favor of the Assembly Language Generation task (ALG, Section \ref{sec:ALG}) and the Similar Function Prediction task (SFP, Section \ref{sec:SFP}).

\begin{figure}[!t]
    \centering
    \includegraphics[width=3in]{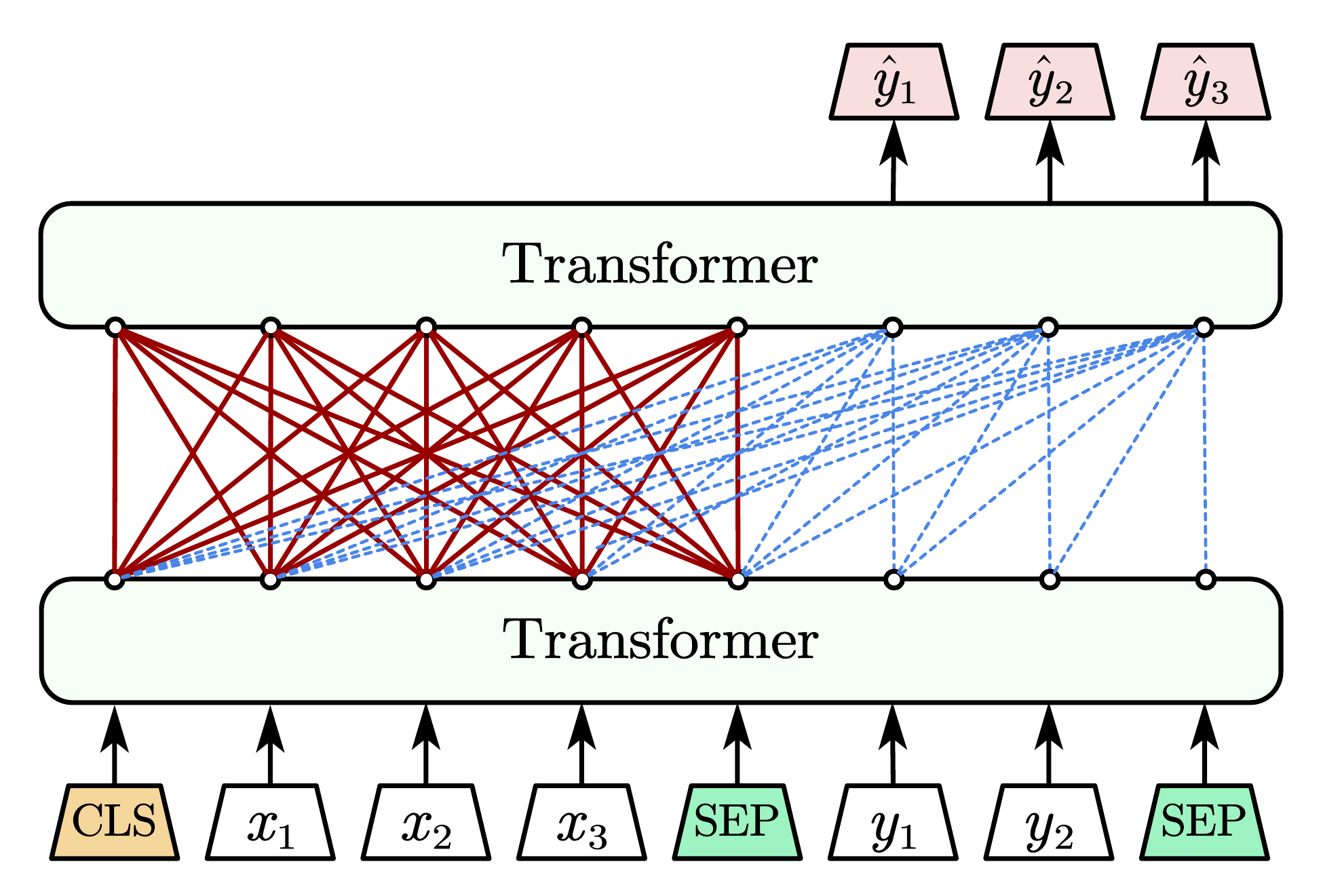}
    \caption{Assembly Language Generation.}
    \label{fig:ALG}
\end{figure}

\subsubsection{Assembly Language Generation}
\label{sec:ALG}

ALG leverages an attention mask matrix to define bidirectional attention and unidirectional attention. As shown in Figure \ref{fig:ALG}, the input sequence contains a pair of similar functions. The first function in the input sequence uses bidirectional attention, while the second function uses unidirectional attention. It allows the model to generate the second function according to the first one.

For the input pair of functions $\mathbf{F}=[ {{x}_{1}},{{x}_{2}},\cdots ,{{x}_{n}} ]$ and ${{\mathbf{F}}^{\prime }}=[ {{y}_{1}},{{y}_{2}},\cdots ,{{y}_{m}} ]$, the input tokens for UniASM are $[ \text{CLS},{{x}_{1}},\cdots ,{{x}_{n}},\text{SEP},{{y}_{1}},\cdots ,{{y}_{m}},\text{SEP} ]$. The goal of ALG is to correctly predict the second function ${{\mathbf{F}}^{\prime }}$ according to the first function $\mathbf{F}$. When we get the predict value ${{\mathbf{\hat{F}}}^{\prime }}=[ {{{\hat{y}}}_{1}},{{{\hat{y}}}_{2}},\cdots ,{{{\hat{y}}}_{m}} ]$, the softmax is applied to the result:

\begin{equation}
    p( {{{\hat{y}}}_{i}}|\mathbf{F} )=\frac{\exp ( {{{\hat{y}}}_{i}} )}{\sum\nolimits_{k=1}^{m}{\exp ( {{{\hat{y}}}_{k}} )}},
\end{equation}

\noindent where ${{\hat{y}}_{i}}$ denotes the predict value of ${{y}_{i}}$. ALG uses cross-entropy to calculate the loss as follows:

\begin{equation}
    \underset{\theta }{\mathop{\min }}\,{{\mathcal{L}}_{ALG}}( \theta  )=\sum\limits_{i}{-\log p( {{{\hat{y}}}_{i}}|\mathbf{F} )}.
\end{equation}

\begin{figure}[!t]
    \centering
    \includegraphics[width=4in]{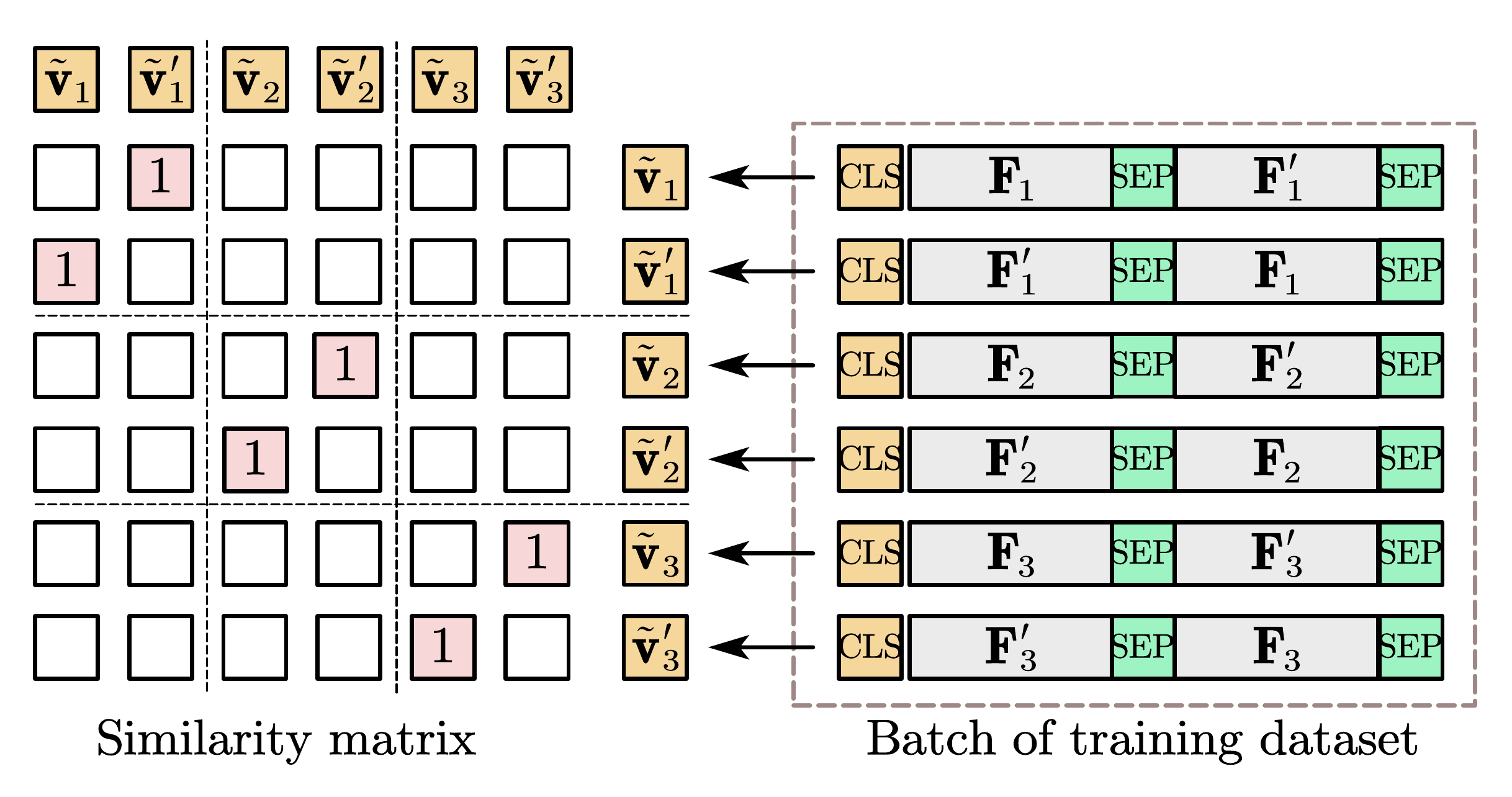}
    \caption{Similar Function Prediction.}
    \label{fig:SFP}
\end{figure}

\subsubsection{Similar Function Prediction}
\label{sec:SFP}

SFP processes one batch rather than a pair of functions at a time. As shown in Figure \ref{fig:SFP}, each sample in the batch is a pair of similar functions, such as $[\text{CLS}]\text{ }\mathbf{F}\text{ }[\text{SEP}]\text{ }{{\mathbf{F}}^{\prime }}\text{ }[\text{SEP}]$, where $\mathbf{F}$ and ${{\mathbf{F}}^{\prime }}$ are similar functions. We swap these two functions to construct a new sample $[\text{CLS}]\text{ }{{\mathbf{F}}^{\prime }}\text{ }[\text{SEP}]\text{ }\mathbf{F}\text{ }[\text{SEP}]$ and place it after the original one. So, each batch should contain an even number of samples.

The embedding of the $k$-th function in the batch is ${{\mathbf{v}}_{k}}=[{{v}_{1}},{{v}_{2}},...,{{v}_{d}}]$, where $d$ is the hidden size. Then the elements in the vector are L2 normalized:

\begin{equation}
    {{\tilde{v}}_{i}}=\frac{{{v}_{i}}}{\sqrt{\sum\nolimits_{j=1}^{d}{v_{j}^{2}}}}.
\end{equation}

The normalized function embedding vector can be obtained: ${{\mathbf{\tilde{v}}}_{k}}=[{{\tilde{v}}_{1}},{{\tilde{v}}_{2}},...,{{\tilde{v}}_{d}}]$. We take all the normalized vector of the batch to construct the embedding matrix $\mathbf{\tilde{V}}={{[{{\mathbf{\tilde{v}}}_{1}},{{\mathbf{\tilde{v}}}_{2}},...,{{\mathbf{\tilde{v}}}_{b}}]}^{\top }}$, where $b$ is the batch size.

To calculate the similarity between two functions in the batch, we dot product the embedding matrix $\mathbf{\tilde{V}}$ with its transposed matrix ${{\mathbf{\tilde{V}}}^{\top }}$:

\begin{equation}
    \mathbf{S}=\mathbf{\tilde{V}}\cdot {{\mathbf{\tilde{V}}}^{\top }}=\{ {{s}_{ij}} \},i,j\in [ 1,2,\cdots ,b ].
\end{equation}

The result $\mathbf{S}$ is called the similarity matrix. Each value in the similarity matrix denotes the similarity of two functions. The idea is based on the fact that the value of the dot product of unit vectors is equal to $\cos ( \varphi  )$, where $\varphi $ denotes the angle between two vectors. The more similar the vectors are, the smaller the angle between them should be. That is, the dot product of vectors of similar functions should be closer to 1, and the dot product of vectors of different functions should be closer to -1.

It should be noted that values on the diagonal in the similarity matrix are all equal to 1 because they are dot products of the same unit vector. However, we only care about the value of the two similar functions. To avoid the effect of the diagonal elements, we set all diagonal elements to negative infinity:

\begin{equation}
    \mathbf{S}=\mathbf{\tilde{V}}\cdot {{\mathbf{\tilde{V}}}^{\top }}-\mathbf{\Lambda }[ +\infty  ],
\end{equation}

\noindent where $\mathbf{\Lambda }[ +\infty  ]$ denotes a diagonal matrix, whose values are set to infinity. Each row of the matrix needs to be processed by softmax layer as:

\begin{equation}
    p( {{s}_{ij}} )=\frac{\exp ( {{s}_{ij}} )}{\sum\nolimits_{k=1}^{b}{\exp ( {{s}_{ik}} )}},
\end{equation}

\noindent where ${{s}_{ij}}$ denotes the similarity of the $i$-th function and the $j$-th function in the batch. SFP uses cross-entropy to calculate the loss as follows:

\begin{equation}
    \underset{\theta }{\mathop{\min }}\,{{\mathcal{L}}_{SFP}}( \theta  )=\sum\limits_{k}{-\log p( {{s}_{ik}} )}.
\end{equation}

The loss function of UniASM is the combination of the two loss functions:

\begin{equation}
    \underset{\theta }{\mathop{\min }}\,\mathcal{L}( \theta  )={{\mathcal{L}}_{ALG}}( \theta  )+{{\mathcal{L}}_{SFP}}( \theta  ).
\end{equation}

\section{Experimental Setups}

\label{sec:setups}
\subsection{Dataset}
\label{sec:dataset}
\subsubsection{Training Dataset}
\label{sec:train dataset}

As shown in Table \ref{tab:train dataset}, we collected seven open-source projects commonly used under Linux as the training dataset for UniASM.

\textbf{Compilation} We used two compilers (GCC-7.5 and Clang-10) with four optimization levels (O0/O1/O2/O3). In addition, the obfuscator Ollvm14 \cite{Junod2015ObfuscatorLLVMS} was used to generate different obfuscated codes (sub/fla/bcf) with the four optimization levels, where “sub” denotes instruction substitution, “fla” denotes control flow flattening, “bcf” denotes bogus control flow. Thus, we got 11 different results for each of the input functions. It should be noted that all source codes were compiled with the option “\textit{-fno-inline}” to avoid function inlining. The main reason is that the function inlining can interfere with similar function pairs, which is detrimental to the training of the model. After the compilation, we obtained 133 binaries for each compilation environment.

\textbf{Disassembly} UniASM is designed to generate embeddings for assembly codes. We disassembled the binaries with the help of Radare2 \cite{Radare2022RADARE2} and saved the functions in separate files. There is a slight difference in the number of functions obtained by disassembling the binary for different compilation options and code obfuscations. We got 12,694 unique functions (GCC-O0) and about 260K disassembly files.

\textbf{Similar function pairs} The training data for UniASM was constructed from similar function pairs. As shown in Table \ref{tab:func pairs}, we combined some of the different disassembly results for each function to form 40 similar function pairs. The numbers in the table indicate the number of function pairs to be generated, “-” means no function pairs are generated. For the same compiler, only the optimization level needs to be considered, and there are six combinations (O0-O1, O0-O2, O0-O3, O1-O2, O1-O3, and O2-O3). For the different compilers, all 16 combinations were considered. For the code obfuscations, we only combined the obfuscated code with the normal code because we expected UniASM to learn the obfuscation features. We obtained about 500K similar function pairs in total.

\textbf{Dataset generation} We generated two sequences for each function pair according to the following steps:

\begin{enumerate}
    \item Small functions with less than ten instructions were filtered to avoid semantically meaningless functions. 
    \item A new function pair was generated by swapping the two functions.
    \item The tokenizer converted each function pair into a token sequence. 
    \item All the sequences were shuffled randomly and divided into two parts: 90\% for training and 10\% for validation.
\end{enumerate}

The training dataset contains 428K sequences, and the validation dataset contains about 47K sequences.

\begin{table}[!t]
\renewcommand{\arraystretch}{1.3}
\caption{Projects Used for Training}
\label{tab:train dataset}
\centering
\begin{tabular}{cccrr}
\hline
Projects & Version & Binaries & Func. & ASMs\\
\hline
Binutils & 2.37 & 16 & 5,465 & 107,098\\
Coreutils & 9.0 & 106 & 2,321 & 47,406\\
Diffutils & 3.8 & 4 & 592 & 12,008\\
Findutils & 4.8.0 & 4 & 898 & 18,135\\
Tcpdump & 4.9.3 & 1 & 1,448 & 32,243\\
Gmp & 6.2.1 & 1 & 760 & 16,777\\
Curl & 7.82.0 & 1 & 1,210 & 26,455\\
\hline
Total & - & 133 & 12,694 & 260,122\\
\hline
\end{tabular}
\end{table}

\begin{table}[!t]
\renewcommand{\arraystretch}{1.3}
\caption{Similar Function Pairs}
\label{tab:func pairs}
\centering
\begin{tabular}{cccccc}
\hline
 & GCC & Clang & Ollvm-sub & Ollvm-fla & Ollvm-bcf\\
\hline
GCC & 6 & 16 & - & - & -\\
Clang & - & 6 & - & - & -\\
Ollvm & - & - & 4 & 4 & 4\\
\hline
\end{tabular}
\end{table}

\subsubsection{Evaluation Dataset}
\label{sec:eval dataset}

This paper prepared three datasets to evaluate our model and the baselines:

\textbf{DS-BinKit} is based on BinKit-2.0 \cite{Donkwan2023Binkit, BinKitDataset}, and used for evaluating the performance of the models (Section \ref{sec:eval perf}, X-COM and X-OPT). BinKit-2.0 pre-compiled 50 projects with 8 architectures, 6 optimization levels (O0/O1/ O2/O3/Os/Ofast), and 18 compilers (Clang4-13, GCC4-11). We select all the x86\_64 binaries with four optimization levels (O0/O1/O2/O3), resulting in 26,458 binaries and 2,710,964 functions.

\textbf{DS-OBF} is generated by seven open-source projects (libarchive-3.1.2, libav-12, libgd-2.1.1, libpcap-1.9.1, libressl-2.7.0, openjpeg-2.1, and openssh-7.3p1), and used for performance evaluation (Section \ref{sec:eval perf}, X-OBF) and ablation studies (Section \ref{sec:eval abla}). All projects are not included in the training dataset and cover different application scenarios. The projects were compiled by two compilers (GCC and Clang) with the four optimization levels (O0/O1/O2/O3) and by Ollvm14 with three obfuscations (sub/fla/bcf), obtaining a total of 176,036 functions.

\textbf{DS-VUL} is a set of vulnerabilities and the affected projects, as shown in Table \ref{tab:perf vuln}, and is used for evaluating performance on real-world vulnerability searching (Section \ref{sec:eval vuln}). We selected eight vulnerabilities from a known vulnerabilities dataset \cite{SecretPatch2022DATA} as the search targets. Then, we compiled the affected projects into 11 variants.

\subsection{Baselines}
\label{sec:baselines}
We compared UniASM to the following six baselines:

\textbf{InnerEye \cite{Zuo2019NeuralMT}} uses LSTM in a Siamese architecture for binary code similarity detection. Specifically, it first leverages word2vec to generate instruction embedding and then feeds them to the Siamese architecture to learn basic block embedding. We obtained function embedding by taking the entire function as input. We used their official open-source code and pre-trained model \cite{InnerEye2022code} with its default parameters for evaluation.

\textbf{Asm2vec \cite{Ding2019Asm2VecBS}} is a PV-DM-based model for assembly language embedding. It uses random walks on the CFG to sample instruction sequences and then uses the PV-DM model to learn the embedding of the assembly language. The original paper of Asm2vec shows that their dataset contains function names of system libraries, but our validation dataset does not contain this kind of information. Asm2vec is not open source. We used an unofficial version \cite{Oalieno2022ASM2VEC} that is publicly available and configured the default parameters for evaluation.

\textbf{SAFE \cite{Massarelli2019SAFESF}} is an Attention-based model for assembly language embedding. It employs an RNN architecture with attention mechanisms to generate function embeddings. We used their official open-source code and pre-trained model \cite{SAFE2022SAFE} with its default parameters for evaluation.

\textbf{PalmTree \cite{Li2021PalmTreeLA}} is a BERT-based model for assembly instruction embedding. It uses three pre-training tasks to learn the characteristics of assembly instructions and generate the instruction embeddings. The evaluation is based on their official open-source code and pre-trained model \cite{PalmTree2022PALM} with its default parameters. Since PalmTree cannot embed function directly, we trained a GAT network to generate the function embeddings. First, we used PalmTree to pre-process our training set and obtained the ACFG of each function (following the authors’ approach of using mean pooling to generate basic block embeddings). Then, we trained a Siamese GAT network to predict the similarity between pairs of ACFGs in the training set. Finally, we used the trained GAT network to generate function embeddings.

\textbf{jTrans \cite{Wang2022jTransJT}} is a jump-aware BERT-based model for assembly language embedding. It retains the jump relationships between instructions when generating input samples for BERT and allows BERT to learn the control flow information of the code. We evaluated jTrans using its best performing fine-tuned model \cite{jTrans2022JTRANS} with all parameters kept at their default values.

\textbf{kTrans \cite{Zhu2023KTrans}} integrates domain knowledge into a Transformer framework for assembly language embedding. It feeds explicit knowledge as additional inputs to the Transformer to model implicit dependencies in assembly language. We evaluated kTrans using its official code and pre-trained model \cite{kTrans2023}.

\subsection{Evaluation Metrics}
\label{sec:eval metric}
The task of function similarity search is often used to measure the performance of BCSD models. The function similarity search aims to find similar functions in a large pool of functions for the input function. The input function is selected from a source function pool, and the model searches the target function pool to find similar functions. The source and target function pools are defined as:

\begin{align}
{{\mathcal{F}}_{src}}&=\{ {{f}_{1}},{{f}_{2}},\cdots ,{{f}_{i}},\cdots ,{{f}_{n}} \} \\ 
{{\mathcal{G}}_{dst}}&=\{ {{g}_{f1}},{{g}_{f2}},\cdots ,{{g}_{fi}},\cdots ,{{g}_{fn}} \}
\end{align}

The source function pool ${{\mathcal{F}}_{src}}$ contains $n$ functions, i.e., the pool size is $n$. Each input function ${{f}_{i}}\in {{\mathcal{F}}_{src}}$ corresponds to a ground truth function ${{g}_{fi}}\in {{\mathcal{G}}_{dst}}$.

The metrics of Mean Reciprocal Rank (MRR) and Recall@$k$ are used to evaluate the performance. We take the top-$k$ results for each query and sort them according to the similarity score. The MRR metric is the average of the reciprocal ranks of results for a sample of the queries, while the Recall@$k$ metric is the ratio of successful queries to the pool size (a successful query means the true ground function is in the top-$k$ results):

\begin{align}
\text{MRR}( {{\mathcal{F}}_{src}} )&=\frac{1}{| {{\mathcal{F}}_{src}} |}\sum\limits_{{{f}_{i}}\in {{\mathcal{F}}_{src}}}{\frac{1}{\text{Rank}( {{g}_{fi}}|{{f}_{i}} )}} \\ 
 \text{Recall}@k( {{\mathcal{F}}_{src}} )&=\frac{1}{| {{\mathcal{F}}_{src}} |}\sum\limits_{{{f}_{i}}\in {{\mathcal{F}}_{src}}}{[\![ \text{Rank}( {{g}_{fi}}|{{f}_{i}} )\le k ]\!]}
\end{align}

\noindent where $\text{Rank}( {{g}_{fi}}|{{f}_{i}} )$ refers to the ranking position of the first hit function for the $i$-th query. $[\![ \cdot  ]\!]$ is an identity function that outputs 1 if the expression inside is evaluated to be true and 0 otherwise.

\subsection{Hyperparameter Selection}
\label{sec:hyperpara}

We chose the following hyperparameters for UniASM: 4 transformer layers, 12 attention heads, max sequence length of 256, vocabulary size of 21000, and intermediate size of 3072. For training, we chose the batch size of 8, and the learning rate is set to 5e-5 with the warm-up of 4 steps.

\section{Evaluation}
\label{sec:eval}

The evaluation aims to answer the following questions:

     RQ1: How accurate is UniASM in BCSD tasks compared with other baselines? (Section \ref{sec:eval perf})
     
     RQ2: What impact do different factors have on the model’s accuracy in BCSD tasks? (Section \ref{sec:eval abla})
     
     RQ3: How effective is UniASM at searching known vulnerabilities? (Section \ref{sec:eval vuln})

All programs were compiled and pre-processed on an Ubuntu 20.04 server with 16GB RAM and Intel 8 core 3.0GHz CPU. In most cases, we used Radare2 for disassembling binary programs to generate assembly code. One of the baseline methods, jTrans, requires IDA pro \cite{Hexrays2022IDA} to disassemble the binary program. We trained UniASM on one TPU v3-8 chip, and all evaluation experiments were run on a workstation with Intel Core i7-13700K CPU, 64GB RAM, and NVIDIA GeForce GTX 1080 Ti 11GB GPU.

\subsection{Performance}
\label{sec:eval perf}

This paper evaluated UniASM and the baselines on three BCSD scenarios: cross-compilers (X-COM), cross-optimization levels (X-OPT), and cross-obfuscations (X-OBF). Table \ref{tab:BCSD perf-xcom.}-\ref{tab:BCSD perf-obf.} shows the scores of MRR and Recall@1 for UniASM and the baselines. The Recall@1 metric captures the ratio of functions correctly matched at the first position of the search results.

We conducted X-COM and X-OPT evaluations on the DS-BinKit dataset and X-OBF on the DS-OBF dataset. DS-BinKit, as a third-party dataset, can improve the fairness of the experiments. However, DS-BinKit only contains binaries generated using different compilers and optimization options, but not binaries generated using obfuscation methods. Therefore, we compiled our own DS-OBF dataset to complete the X-OBF evaluation. To ensure the fairness of the experiments, the projects and binaries used in DS-OBF did not appear in our training set. Additionally, the experimental results in \cite{Wang2022jTransJT} indicate that the choice of pool size has a significant impact on the evaluation score. Therefore, we also selected two pool sizes, 100 and 10,000, to evaluate the performance of UniASM and baseline methods in varying degrees of difficulty.

For X-COM, we classified the functions in DS-BinKit according to the type of compiler, resulting in 18 function pools (10 Clang compilers and 8 GCC compilers). We selected six compiler pairs with the greatest differences from these 18 function pools for experimentation: Clang4-Clang13, Clang5-Clang12, GCC4-GCC11, GCC5-GCC10, Clang4-GCC11, and Clang13-GCC4 to verify the model's performance in cross-compiler similarity comparison. 

For X-OPT, we classified the functions in DS-BinKit according to four compilation optimization options (O0/O1/O2/O3), and conducted similarity search experiments on all possible pairs: O0-O1, O0-O2, O0-O3, O1-O2, O1-O3, and O2-O3. 

For X-OBF, we classified the functions in DS-OBF into four categories based on code obfuscation methods: none, bcf, fla, and sub, and then paired them to form six similarity search experiments: none-bcf, none-fla, none-sub, bcf-fla, bcf-sub, and fla-sub.

\begin{table*}[!t]
\renewcommand{\arraystretch}{1}
\caption{BCSD Performance of X-COM on DS-BinKit}
\label{tab:BCSD perf-xcom.}
\centering
\begin{threeparttable}
\small
\begin{tabular}{cl@{\extracolsep{4pt}}c@{\extracolsep{4pt}}c@{\extracolsep{4pt}}c@{\extracolsep{4pt}}c@{\extracolsep{4pt}}c@{\extracolsep{4pt}}c@{\extracolsep{4pt}}c}
\hline
 & \multicolumn{1}{c}{\multirow{2}{*}{Models}} & \multicolumn{6}{c}{Recall@1} & \multirow{2}{*}{Avg.} \\
 & \multicolumn{1}{c}{} & C4-C13 & C5-C12 & G4-G11 & G5-G10 & C4-G11 & C13-G4 & \\ \cline{1-9} 
 
\multirow{6}{*}{\rotatebox{90}{pool=100}} 
 & InnerEye     & .500 & .540 & .490 & .570 & .230 & .840 & .528 \\
 & Asm2Vec      & .770 & .840 & .860 & .840 & .540 & .940 & .798 \\
 & SAFE         & .860 & .920 & .860 & .910 & .810 & .940 & .883 \\
 & Palmtree     & .810 & .890 & .860 & .890 & .570 & .940 & .827 \\
 & kTrans       & .910 & .930 & \textbf{.930} & \textbf{.930} & .880 & .960 & \textbf{.923} \\
 & jTrans       & \textbf{.920} & .910 & .850 & .900 & .690 & \textbf{.970} & .873 \\
 & UniASM       & \textbf{.920} & \textbf{.940} & .900 & .910 & \textbf{.920} & .950 & \textbf{.923} \\ \cline{1-9}
  
\multirow{6}{*}{\rotatebox{90}{pool=10000}} 
 & InnerEye     & .098 & .111 & .128 & .171 & .017 & .280 & .134 \\
 & Asm2Vec      & .222 & .245 & .263 & .271 & .074 & .340 & .236 \\
 & SAFE         & .285 & .318 & .329 & .338 & .179 & .401 & .308 \\
 & Palmtree     & .300 & .335 & .331 & .337 & .111 & .397 & .302 \\
 & kTrans       & .328 & .360 & \textbf{.367} & \textbf{.373} & .245 & .408 & .347 \\ 
 & jTrans       & .317 & .340 & .297 & .368 & .152 & \textbf{.409} & .314 \\
 & UniASM       & \textbf{.329} & \textbf{.363} & .360 & .367 & \textbf{.299} & .407 & \textbf{.354} \\ \hline

 & \multicolumn{1}{c}{\multirow{2}{*}{Models}} & \multicolumn{6}{c}{MRR} & \multirow{2}{*}{Avg.} \\
 & \multicolumn{1}{c}{} & C4-C13 & C5-C12 & G4-G11 & G5-G10 & C4-G11 & C13-G4 & \\ \cline{1-9} 
\multirow{6}{*}{\rotatebox{90}{pool=100}} 
 & InnerEye     & .602 & .633 & .596 & .655 & .329 & .886 & .617 \\
 & Asm2Vec      & .842 & .892 & .900 & .873 & .651 & .957 & .852 \\
 & SAFE         & .914 & .947 & .893 & .933 & .879 & .960 & .921 \\
 & Palmtree     & .864 & .924 & .886 & .912 & .690 & .961 & .873 \\
 & kTrans       & .952 & .963 & \textbf{.943} & \textbf{.943} & .925 & .978 & .951 \\
 & jTrans       & \textbf{.953} & .950 & .899 & .929 & .754 & \textbf{.983} & .911 \\
 & UniASM       & .952 & \textbf{.968} & .932 & .941 & \textbf{.952} & .968 & \textbf{.952}  \\ \cline{1-9}
  
\multirow{6}{*}{\rotatebox{90}{pool=10000}} 
 & InnerEye     & .161 & .181 & .191 & .248 & .032 & .403 & .202 \\
 & Asm2Vec      & .337 & .367 & .381 & .391 & .127 & .479 & .347 \\
 & SAFE         & .421 & .462 & .470 & .480 & .289 & .555 & .446 \\
 & Palmtree     & .439 & .480 & .468 & .477 & .187 & .552 & .434 \\
 & kTrans       & .476 & .512 & \textbf{.516} & \textbf{.523} & .373 & .564 & .494 \\
 & jTrans       & .462 & .487 & .429 & .518 & .244 & \textbf{.565} & .451 \\
 & UniASM       & \textbf{.478} & \textbf{.517} & .509 & .517 & \textbf{.443} & .563 & \textbf{.504}  \\ \hline
\end{tabular}

\end{threeparttable}
\end{table*}

\begin{table*}[!t]
\renewcommand{\arraystretch}{1}
\caption{BCSD Performance of X-OPT on DS-BinKit}
\label{tab:BCSD perf-xopt.}
\centering
\begin{threeparttable}
\small

\begin{tabular}{cl@{\extracolsep{4pt}}c@{\extracolsep{4pt}}c@{\extracolsep{4pt}}c@{\extracolsep{4pt}}c@{\extracolsep{4pt}}c@{\extracolsep{4pt}}c@{\extracolsep{4pt}}c}
\hline
 & \multicolumn{1}{c}{\multirow{2}{*}{Models}} & \multicolumn{6}{c}{Recall@1} & \multirow{2}{*}{Avg.} \\
 & \multicolumn{1}{c}{} & O0-O1 & O0-O2 & O0-O3 & O1-O2 & O1-O3 & O2-O3 & \\ \cline{1-9} 
\multirow{6}{*}{\rotatebox{90}{pool=100}} 
 & InnerEye     & .190 & .140 & .130 & .450 & .410 & .700 & .337 \\
 & Asm2Vec      & .240 & .370 & .240 & .730 & .680 & .800 & .510 \\
 & SAFE         & .636 & .556 & .525 & .768 & .687 & .768 & .657 \\
 & Palmtree     & .250 & .220 & .220 & .710 & .620 & .780 & .467 \\
 & kTrans       & .330 & .320 & .310 & .800 & .730 & .830 & .553 \\
 & jTrans       & .660 & .570 & .550 & \textbf{.810} & \textbf{.770} & \textbf{.850} & .702 \\
 & UniASM       & \textbf{.800} & \textbf{.750} & \textbf{.720} & .800 & .740 & .800 & \textbf{.768} \\ \cline{1-9}
  
\multirow{6}{*}{\rotatebox{90}{pool=10000}} 
 & InnerEye     & .004 & .004 & .004 & .118 & .109 & .341 & .097 \\
 & Asm2Vec      & .023 & .016 & .016 & .234 & .217 & .334 & .140 \\
 & SAFE         & .087 & .064 & .060 & .252 & .230 & .403 & .183 \\
 & Palmtree     & .013 & .009 & .008 & .258 & .236 & .389 & .152 \\
 & kTrans       & .031 & .025 & .023 & .319 & .301 & .431 & .188 \\
 & jTrans       & .136 & .116 & .108 & \textbf{.345} & \textbf{.327} & \textbf{.455} & .248 \\
 & UniASM       & \textbf{.211} & \textbf{.186} & \textbf{.176} & .320 & .299 & .418 & \textbf{.269} \\ \hline

 & \multicolumn{1}{c}{\multirow{2}{*}{Models}} & \multicolumn{6}{c}{MRR} & \multirow{2}{*}{Avg.} \\
 & \multicolumn{1}{c}{} & O0-O1 & O0-O2 & O0-O3 & O1-O2 & O1-O3 & O2-O3 & \\ \cline{1-9} 
\multirow{6}{*}{\rotatebox{90}{pool=100}} 
 & InnerEye     & .295 & .267 & .250 & .546 & .501 & .767 & .438 \\
 & Asm2Vec      & .399 & .466 & .346 & .817 & .760 & .860 & .608 \\
 & SAFE         & .770 & .697 & .652 & .847 & .770 & .835 & .762 \\
 & Palmtree     & .365 & .343 & .337 & .782 & .711 & .828 & .561 \\
 & kTrans       & .475 & .466 & .449 & .868 & .803 & .888 & .658 \\
 & jTrans       & .793 & .719 & .699 & \textbf{.885} & \textbf{.844} & \textbf{.905} & .807 \\
 & UniASM       & \textbf{.892} & \textbf{.850} & \textbf{.816} & .881 & .832 & .868 & \textbf{.856} \\ \cline{1-9}
  
\multirow{6}{*}{\rotatebox{90}{pool=10000}} 
 & InnerEye     & .011 & .010 & .010 & .174 & .161 & .450 & .136 \\
 & Asm2Vec      & .044 & .034 & .032 & .335 & .312 & .458 & .203 \\
 & SAFE         & .160 & .123 & .114 & .366 & .338 & .529 & .272 \\
 & Palmtree     & .026 & .020 & .019 & .366 & .338 & .517 & .214 \\
 & kTrans       & .060 & .051 & .048 & .447 & .426 & .564 & .266 \\
 & jTrans       & .215 & .191 & .182 & \textbf{.471} & \textbf{.451} & \textbf{.586} & .349 \\
 & UniASM       & \textbf{.336} & \textbf{.301} & \textbf{.286} & .452 & .426 & .550 & \textbf{.392} \\ \hline
\end{tabular}

\end{threeparttable}
\end{table*}

\begin{table*}[!t]
\renewcommand{\arraystretch}{1}
\caption{BCSD Performance of X-OBF on DS-OBF}
\label{tab:BCSD perf-obf.}
\centering
\begin{threeparttable}
\small

\begin{tabular}{cl@{\extracolsep{4pt}}c@{\extracolsep{4pt}}c@{\extracolsep{4pt}}c@{\extracolsep{4pt}}c@{\extracolsep{4pt}}c@{\extracolsep{4pt}}c@{\extracolsep{4pt}}c}
\hline
 & \multicolumn{1}{c}{\multirow{2}{*}{Models}} & \multicolumn{6}{c}{Recall@1} & \multirow{2}{*}{Avg.} \\
 & \multicolumn{1}{c}{} & none-bcf & none-fla & none-sub & bcf-fla & bcf-sub & fla-sub & \\ \cline{1-9} 
\multirow{6}{*}{\rotatebox{90}{pool=100}} 
 & InnerEye     & .440 & .440 & .590 & .420 & .480 & .440 & .468 \\
 & Asm2Vec      & .566 & .707 & .818 & .626 & .667 & .697 & .680 \\
 & SAFE         & .636 & .667 & .909 & .677 & .636 & .576 & .684 \\
 & Palmtree     & .510 & .530 & .780 & .460 & .500 & .480 & .543 \\
 & kTrans       & .615 & .635 & .885 & .531 & .594 & .635 & .649 \\
 & jTrans       & .833 & .897 & \textbf{.987} & .872 & .846 & .897 & .889 \\
 & UniASM       & \textbf{.950} & \textbf{.990} & .960 & \textbf{.960} & \textbf{.920} & \textbf{.970} & \textbf{.958} \\ \cline{1-9}
  
\multirow{6}{*}{\rotatebox{90}{pool=10000}} 
 & InnerEye     & .253 & .259 & .323 & .281 & .286 & .307 & .285 \\
 & Asm2Vec      & .275 & .329 & .439 & .282 & .296 & .357 & .330 \\
 & SAFE         & .347 & .351 & .692 & .333 & .337 & .353 & .402 \\
 & Palmtree     & .305 & .335 & .591 & .282 & .286 & .315 & .352 \\
 & kTrans       & .404 & .384 & .761 & .321 & .379 & .372 & .437 \\
 & jTrans       & .577 & .655 & .784 & .565 & .530 & .576 & .614 \\
 & UniASM       & \textbf{.665} & \textbf{.775} & \textbf{.869} & \textbf{.731} & \textbf{.701} & \textbf{.764} & \textbf{.751} \\ \hline

  & \multicolumn{1}{c}{\multirow{2}{*}{Models}} & \multicolumn{6}{c}{MRR} & \multirow{2}{*}{Avg.} \\
 & \multicolumn{1}{c}{} & none-bcf & none-fla & none-sub & bcf-fla & bcf-sub & fla-sub & \\ \cline{1-9} 
\multirow{6}{*}{\rotatebox{90}{pool=100}} 
 & InnerEye     & .475 & .495 & .666 & .477 & .525 & .498 & .523 \\
 & Asm2Vec      & .600 & .774 & .849 & .665 & .705 & .759 & .725 \\
 & SAFE         & .690 & .732 & .943 & .741 & .687 & .658 & .742 \\
 & Palmtree     & .552 & .565 & .808 & .532 & .544 & .527 & .588 \\
 & kTrans       & .676 & .689 & .901 & .575 & .630 & .681 & .692 \\
 & jTrans       & .876 & .919 & \textbf{.994} & .898 & .888 & .922 & .916 \\
 & UniASM       & \textbf{.964} & \textbf{.990} & .968 & \textbf{.977} & \textbf{.946} & \textbf{.977} & \textbf{.970} \\ \cline{1-9}
  
\multirow{6}{*}{\rotatebox{90}{pool=10000}} 
 & InnerEye     & .267 & .277 & .350 & .291 & .296 & .318 & .300 \\
 & Asm2Vec      & .309 & .384 & .512 & .320 & .337 & .413 & .379 \\
 & SAFE         & .374 & .380 & .736 & .367 & .361 & .375 & .432 \\
 & Palmtree     & .323 & .359 & .634 & .299 & .302 & .332 & .375 \\
 & kTrans       & .432 & .408 & .791 & .336 & .404 & .391 & .460 \\
 & jTrans       & .622 & .709 & .825 & .613 & .575 & .628 & .662 \\
 & UniASM       & \textbf{.733} & \textbf{.834} & \textbf{.902} & \textbf{.792} & \textbf{.757} & \textbf{.816} & \textbf{.806}  \\ \hline
\end{tabular}

\end{threeparttable}
\end{table*}

Table \ref{tab:BCSD perf-xcom.}-\ref{tab:BCSD perf-obf.} presents the MRR and Recall@1 scores of the models in similarity search tasks. UniASM achieved the highest average scores across all tasks. For the more challenging tasks with a pool size of 10,000, UniASM’s average Recall@1 scores in the X-COM, X-OPT, and X-OBF tasks were 0.354, 0.269, and 0.751, respectively, which are 12.7\%, 8.5\%, and 22.3\% higher than the closest baseline method (jTrans). Further analysis of the results revealed that UniASM did better for difficult tasks, such as O0-O3, where UniASM's average Recall@1 score is 57.1\% higher than the closest competitor.

The experimental results demonstrate the effectiveness of UniASM in various BCSD tasks. It is worth noting that UniASM is more lightweight than jTrans (4 transformer layers compared to 12 transformer layers) and has a much smaller training set (260,122 functions compared to 21,085,338 functions). However, UniASM performs better on both the DS-BinKit and DS-OBF datasets, further demonstrating the effectiveness of our backbone network, training tasks, and binary code representation methods.

\subsection{Ablation Studies}
\label{sec:eval abla}

In the ablation studies, we try to find the factors that affect the model’s performance on BCSD. We evaluated the performance of different backbone models (Section \ref{sec:abla models}), training tasks (Section \ref{sec:abla tasks}), instruction normalization (Section \ref{sec:abla norm}), tokenization algorithms (Section \ref{sec:abla token}), function serialization methods (Section \ref{sec:abla corpora}), and max sequence lengths (Section \ref{sec:abla seqlen}) on three typical BCSD tasks: X-OPT (GCC-O0 vs. GCC-O3), X-COM (GCC-O3 vs. Clang-O3), and X-OBF (Ollvm-none vs. Ollvm-bcf). In addition, we compared the embedding space layout of different models to demonstrate their differences visually (Section \ref{sec:abla space}).

We used the Recall@k metric to evaluate the performance of the models. To ensure the fairness of the experiments, we configured all the models with the same hyperparameters (Section \ref{sec:hyperpara}) and used the same training and evaluation datasets (Section \ref{sec:dataset}). The evaluation dataset is DS-OBF, which is a collection of 1,000 functions from different programs.

\subsubsection{Backbone Models}
\label{sec:abla models}

\begin{figure}[!t]
    \centering
    \includegraphics[width=4in]{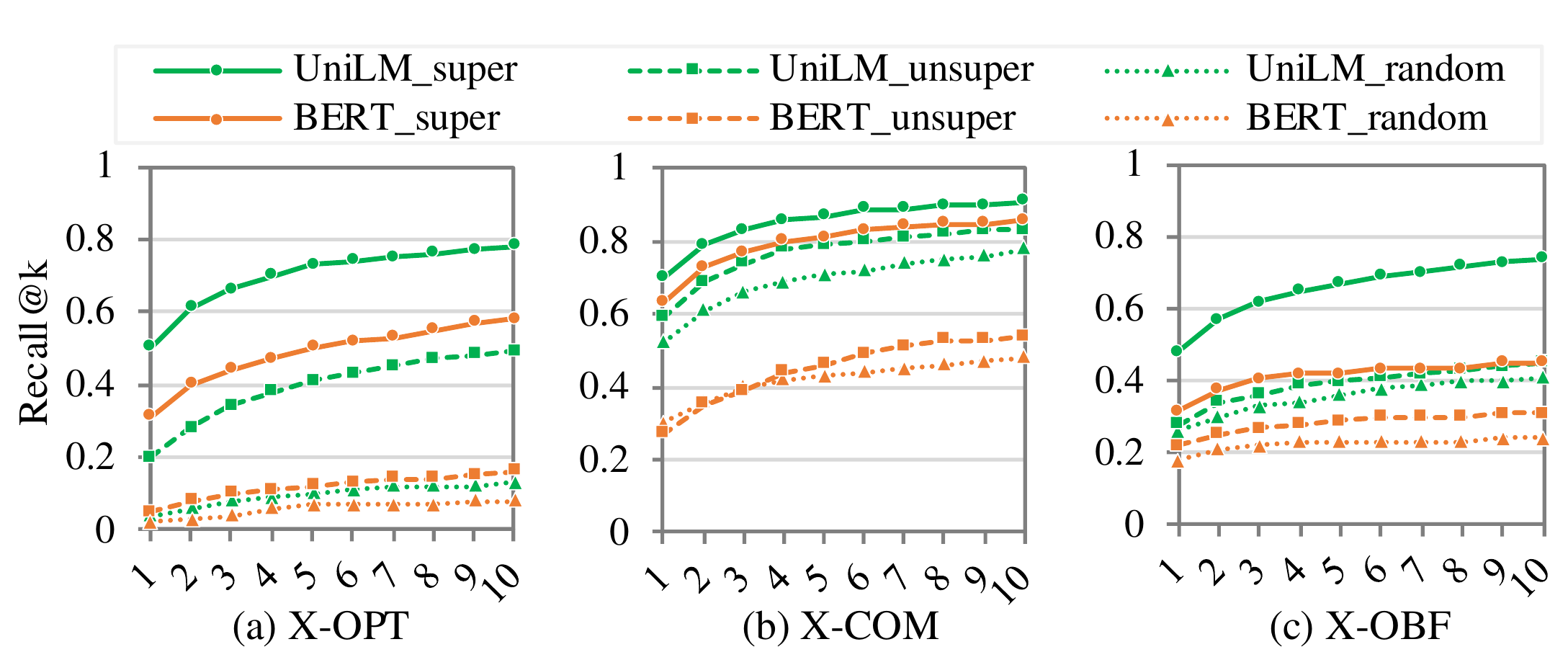}
    \caption{Performance of different models.}
    \label{fig:perf models}
\end{figure}

UniASM is based on the UniLM model, and the BERT model is used as a competitor, which is used in PalmTree and jTrans. Since BCSD take the embeddings of binary code for searching task, in this study, we evaluated the performance of embeddings generated by the two models in three BCSD tasks. To demonstrate the differences between the two models more comprehensively, we designed three different training scenarios:

\begin{enumerate}

\item \textbf{Random parameters} We evaluated BERT and UniLM with the initial random parameters (“bert\_random” and “unilm\_random” in Figure \ref{fig:perf models}) to figure out the baseline performance of the models. 

\item \textbf{Unsupervised learning} We applied only MLM to implement unsupervised training of the two models (“bert\_unsuper” and “unilm\_unsuper” in Figure \ref{fig:perf models}). MLM is BERT’s default training task, which randomly masks 15\% of the tokens in each sequence and trains the model to predict the missing words based on the context.

\item \textbf{Supervised learning} SFP was applied to implement supervised training of the two models (“bert\_super” and “unilm\_super” in Figure \ref{fig:perf models}). SFP is designed to train the model to determine whether two functions are similar.

\end{enumerate}

The results show that UniLM performs much better than BERT in all the BCSD tasks. Both supervised and unsupervised training can improve the performance of the models in the BCSD tasks. We are surprised that UniLM, even with randomly initialized parameters, outperforms the unsupervised trained BERT model in both X-COM and X-OBF tasks.

\subsubsection{Training tasks}
\label{sec:abla tasks}

In this study, we first trained UniASM with one training task at a time, resulting in three pre-trained models: MLM, ALG, and SFP. In addition, we evaluated the combinations of the training tasks, resulting in another two pre-trained models: MLM+SFP and ALG+SFP. We have not tested the combination of MLM and ALG due to the fact that both of them are masking methods with different strategies: MLM implements random position masking, while ALG implements masking of the second half of the sentence.

\begin{figure}[!t]
    \centering
    \includegraphics[width=4in]{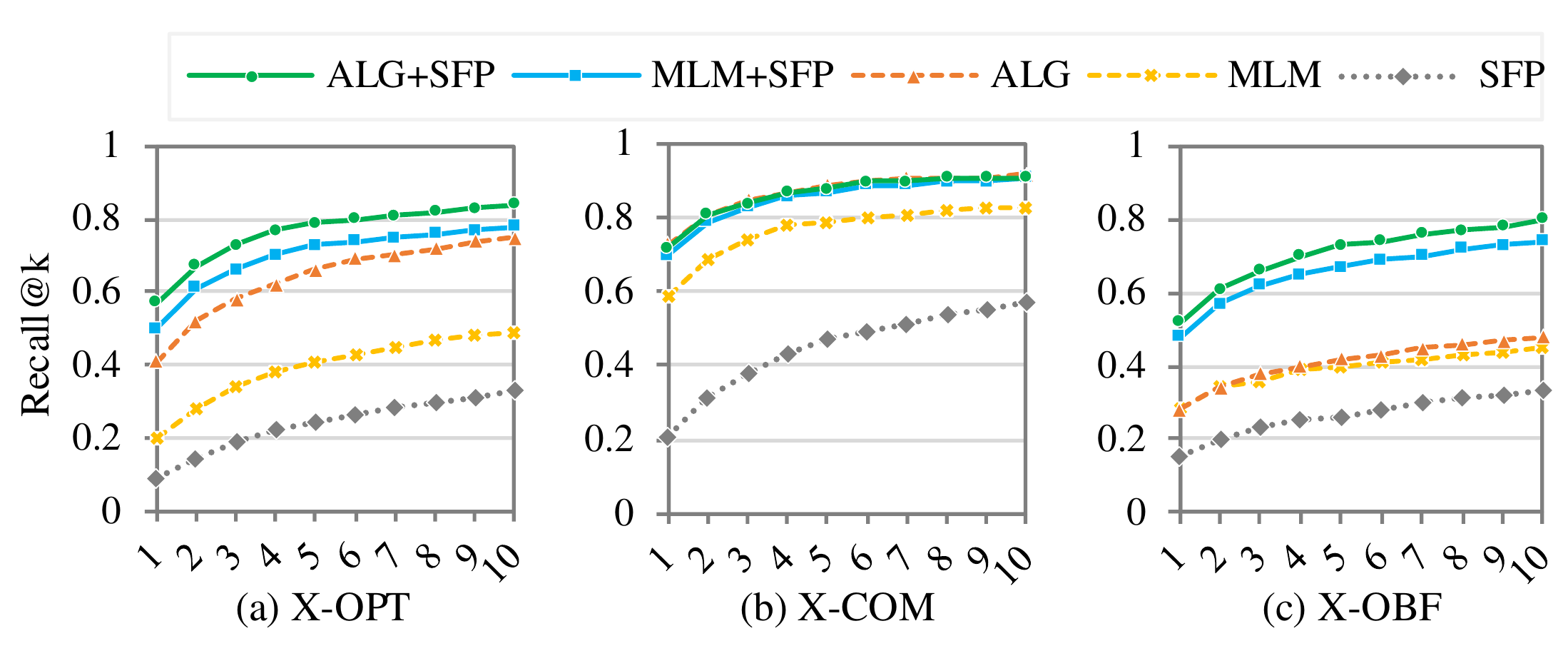}
    \caption{Performance of different training tasks.}
    \label{fig:perf tasks}
\end{figure}

As shown in Figure \ref{fig:perf tasks}, the model with ALG+SFP tasks outperforms all other competitors. When observing the resulting data closely, we find some interesting details:

\begin{itemize}
    \item One is that ALG shows high performance in both X-OPT and X-COM. However, the performance in X-OBF is relatively poor. As shown in Figure \ref{fig:perf tasks}, ALG improves performance over MLM by an average of 66\% in X-OPT and 14\% in X-COM, but only 5\% in X-OBF.
    \item Another is that SFP performs poorly in all BCSD tasks. However, it can significantly improve the model’s performance when combined with MLM or ALG. For the X-OBF task, SFP improves the average Recall@k of MLM from 0.39 to 0.66 and the average Recall@k of ALG from 0.41 to 0.71.
\end{itemize}

The experimental results indicate that ALG is more suitable for BCSD than MLM. One possible reason is that ALG makes the model more focused on the overall semantics of the whole function, while MLM aims to find the missing instructions.

\subsubsection{Instruction Normalization}
\label{sec:abla norm}

Instruction normalization is often used to remove noise in instructions, and the benefits are to reduce the vocabulary size and alleviate the Out-of-Vocabulary (OOV) problem. However, a too-coarse-grained normalization loses a considerable amount of semantic information. A too-fine-grained normalization raises an OOV issue due to many unseen tokens. This study evaluated three typical normalization approaches (KeepI/StripI/NormR) and compared them with our method (Our). We also tested whether keeping some immediate numbers would bring performance improvement (Our*). All experiments were given the same tokenization algorithm, which treats an instruction as a token. For fairness, the vocabulary size is set to the required size for each method during training. The normalization methods are detailed as follows:

\begin{enumerate}
    \item \textbf{Our approach (Our)} Our method keeps all the register names and categorizes the addressing types, detailed in Section \ref{sec:inst rnorm}.
    \item \textbf{Our approach with immediate (Our*)} On top of the default approach, we extract immediate numbers between -4096 and 4096 from the instructions and treat their absolute values as separate tokens. This approach allows us to control the vocabulary growth to only 4097 and not increase the OOV issues.
    \item \textbf{Keep-Immediate (KeepI)} KeepI is a typical fine-grained normalization method that keeps immediate numbers in the instructions. Since the range of immediate numbers is too large, a more practical approach is to retain numbers within a specified range. In this experiment, we kept values between -5000 and 5000 as SAFE did. However, KeepI faces a severe OOV problem due to many unseen tokens.
    \item \textbf{Strip-Immediate (StripI)} StripI is widely used in existing studies \cite{Zuo2019NeuralMT, Ding2019Asm2VecBS, Wang2022jTransJT} to normalize instructions by stripping immediate numbers. This method can avoid interference from different numbers and reduce the vocabulary size. However, StripI does not normalize complex addressing operations, resulting in a more severe OOV problem than our approach.
    \item \textbf{Normalize-Register (NormR)} NormR is a coarse-grained normalization method used in DeepSemantic and DeepBinDiff. Its main idea is to categorize registers by their size and purpose. NormR also further normalizes instructions by stripping immediate numbers. All the strategies make NormR have the smallest vocabulary.
\end{enumerate}

\begin{figure}[!t]
    \centering
    \includegraphics[width=4in]{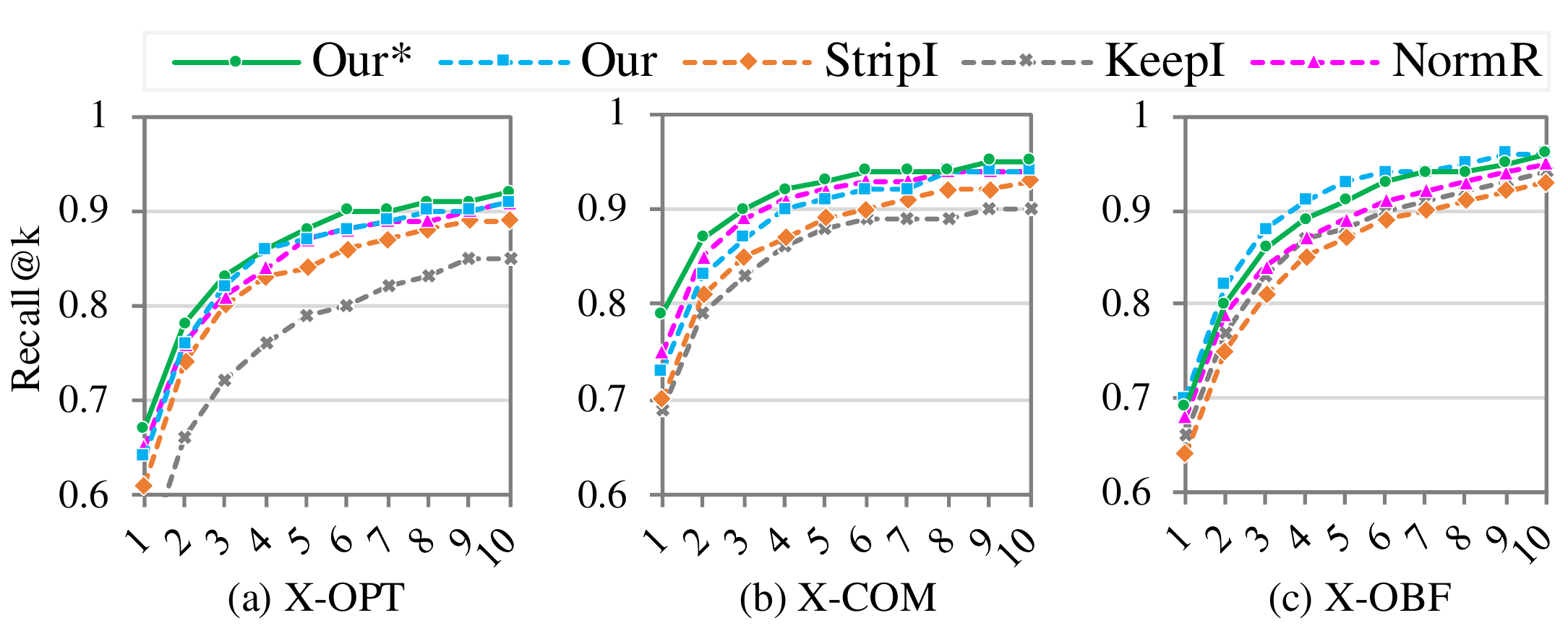}
    \caption{Performance of different normalization methods.}
    \label{fig:perf norm}
\end{figure}

Experimental results (Table \ref{tab:abla norm} and Figure \ref{fig:perf norm}) show that our normalization methods have the least OOV occurrences on DS-OBF and achieve relatively good results in the BCSD tasks. It confirms that there is a negative correlation between performance and OOV. Fine-grained methods, KeepI and StripI, have more OOV occurrences and worse performance. Coarse-grained methods, NormR and our methods, have fewer OOV occurrences and better performance. However, it is necessary to make a balance. NormR loses too much register information, resulting in a small vocabulary but no significant improvement in performance. Besides, when we retain some immediate numbers on top of our method, it improves performance. Overall, our approach is simple and effective. It can maintain rich instruction semantics with a small number of OOV occurrences.

\begin{table}[!t]
\renewcommand{\arraystretch}{1.3}
\caption{Comparison of normalization methods}
\label{tab:abla norm}
\centering
\small
\begin{tabular}{ccrrc}
\hline
\multirow{2}{*}{Methods} & Example & \multirow{2}{*}{Vocab. Size} & \multirow{2}{*}{OOV} & \multirow{2}{*}{Avg. Recall} \\
 & mov eax, [rbp+0x10] & & & \\
\hline
KeepI & mov\_eax,\_[rbp+0x10] & 333,088 & 4,002 & .82 \\
StripI & mov\_eax\_[rbp+NUM] & 73,320 & 502 & .85 \\
NormR & mov\_reg4\_[bp8+NUM] & 6,935 & 176 & .87 \\
\hline
Our & mov\_eax\_SBP & 16,384 & 142 & .88 \\
Our* & mov\_eax\_SBP 0x10 & 20,575 & 142 & .89 \\
\hline
\end{tabular}
\end{table}

\subsubsection{Tokenization Algorithms}
\label{sec:abla token}

Tokenization is an important step in data preprocessing, which converts the normalized instructions to tokens. In this study, we evaluated three tokenization methods (Full-Instruction, Half-Instruction, and Piece-Instruction) designed for assembly code and two tokenization methods (Byte-Pair Encoding \cite{SennrichHB16aBPE} and Word-Piece \cite{SchusterN12WP}) designed for natural language.

\begin{enumerate}
    \item \textbf{Full-Instruction (Full-I, our approach)} Full-I takes a single instruction as a token, as detailed in Section \ref{sec:norm token}. Due to the limitation of input size by the backbone model, coarse-grained tokenization means that more instructions can be used to represent learning. The disadvantage is that it leads to a larger vocabulary and is more prone to the OOV problem.
    \item \textbf{Half-Instruction (Half-I)} Half-I splits each instruction into two parts: the opcode and the operands. This approach can reduce the vocabulary while preserving the semantic information of the operands. However, the sequence size of a function may be twice of Full-I, which may cause the input length to exceed the limit of the model.
    \item \textbf{Piece-Instruction (Piece-I)} Piece-I is used by jTrans, DeepBinDiff, and PalmTree. It is more fine-grained than Full-I and Half-I. Each word in the instruction was treated as a token instead of the whole instruction. Piece-I is easy to implement and can effectively alleviate the OOV problem. However, it destroys the integrity of instructions and increases the difficulty of representation learning. What’s worse, it increases the serialization length of the functions, leading to a higher truncation rate.
    \item \textbf{Byte-Pair Encoding (BPE)} BPE relies on a pre-tokenizer that splits the sentence into words. Since BPE is designed for natural language, we concatenate a function’s instructions into a sentence, separated by spaces. According to this, BPE can easily encode a function into a token sequence.
    \item \textbf{Word-Piece (WP)} WP is the sub-word tokenization algorithm used for BERT, similar to BPE. WP only differs slightly in its symbol pair selection strategy compared to BPE. We prepare the sentences in the same way as for BPE and train WP based on them.
\end{enumerate}

\begin{figure}[!t]
    \centering
    \includegraphics[width=4in]{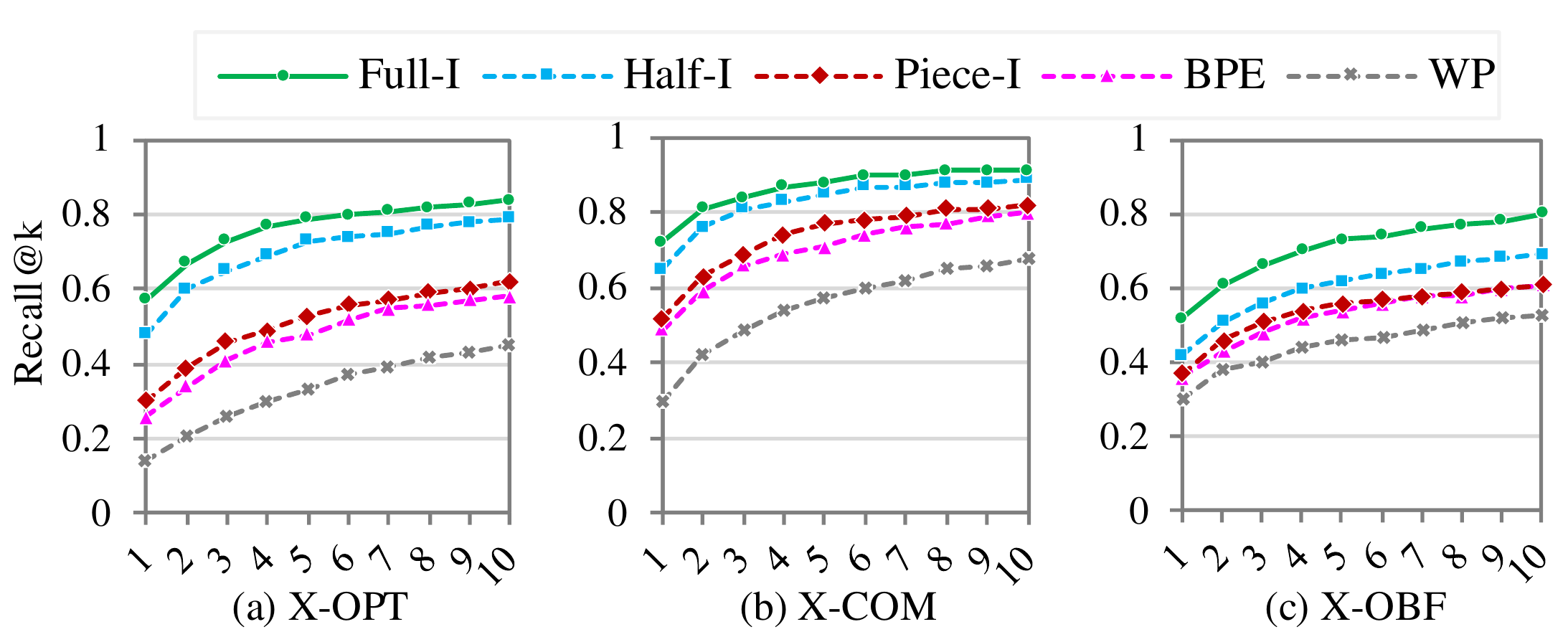}
    \caption{Performance of different tokenization algorithms.}
    \label{fig:perf token}
\end{figure}

The evaluation results (Figure \ref{fig:perf token}) show that Full-I outperforms all other methods. According to the results, we find that coarse-grained tokenization can achieve better performance. Since BPE and WP are all fine-grained tokenization methods, they perform poorly, as expected. The experimental results also indicate that the OOV problem may not be the main factor affecting the embedding performance, and the semantics in the sequence plays a more critical role. However, Full-I requires a vocabulary of over 20,000, while Half-I and Piece-I only need about 7,000 and 4,000, respectively.

\subsubsection{Function Serialization Methods}
\label{sec:abla corpora}

Function serialization aims to serialize a function to a sequence, which can then be tokenized and fed to the NLP model. In this study, we prepared three different serialization methods and also tested the inlining compilation:

\begin{enumerate}
    \item \textbf{Linear (Our approach)} The assembly function in the training dataset was compiled with the no-inlining option “\textit{-fno-inline},” and the function was serialized in linear order (the address order). The linear approach can make the generated sequence contain as many instructions as possible.
    \item \textbf{Random-walk} Random-walk, as used in Asm2Vec, chooses a random path on the CFG of a function, which can extract structure information of the function. The disadvantages are that the randomness leads to the instability of the sequence content, and extracting one execution path of the function shortens the generated sequence’s length.
    \item \textbf{Longest-walk} Longest-walk is an optimized version of random-walk, which chooses the longest path on the CFG of a function. A longer path contains more semantic information about a function. However, it still faces the same problem as random-walk: it can only extract one execution path of the function, which loses some semantics of the function.
    \item \textbf{Inlining compilation} Inlining is the default feature of the compiler, which eliminates call-linkage overhead and can expose other optimization opportunities. Since the evaluation dataset was compiled with default options (inlining turn-on), we designed this experiment to determine whether training with the inlining functions will lead to better performance. However, the new training dataset has 32\% fewer functions than before because some functions are inlined into another function.
\end{enumerate}

\begin{figure}[!t]
    \centering
    \includegraphics[width=4in]{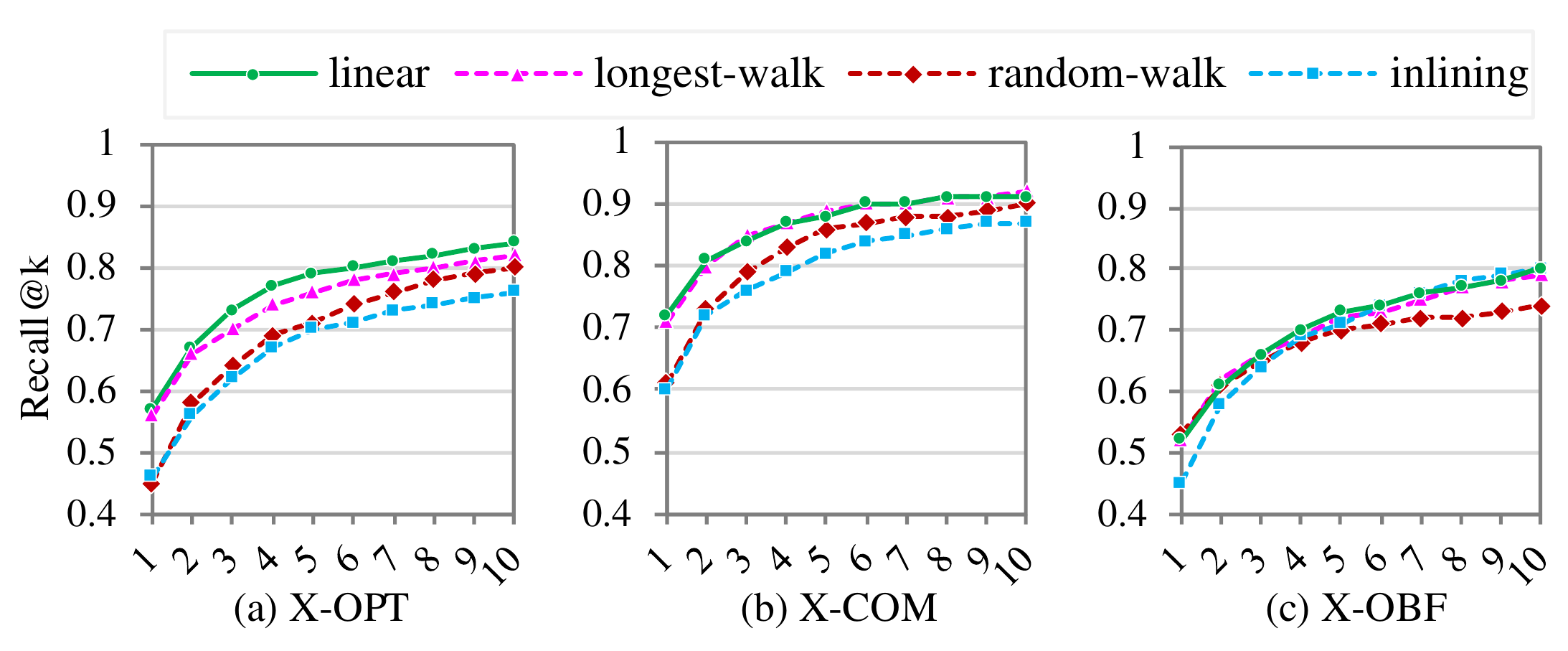}
    \caption{Performance of different function serialization methods.}
    \label{fig:perf corp}
\end{figure}

Surprisingly, the path-based methods (random-walk and longest-walk) did not outperform the default linear method, as shown in Figure \ref{fig:perf corp}. Regarding this phenomenon, we believe that there are some possible explanations: 

\begin{itemize}
    \item First, the sequence generated by the linear order can contain more instructions. According to our statistics, the average instruction count of the linear method is 148, while it is 39 and 62 for random-walk and longest-walk.
    \item Second, the linear order already largely implies the execution order of the function because the instructions in the same basic block are in the right position.
    \item Third, the deep learning model can somewhat adapt to the different orders of basic blocks. The control flow instructions, such as \textit{jmp}, can give a hint to the model.
\end{itemize}

We also find that the inlining dataset is slightly worse than the default dataset. The possible reason is that function inlining reduces the similarity of the function pairs, which makes the training more difficult.

\subsubsection{Max Sequence Lengths}
\label{sec:abla seqlen}

The max sequence length (MaxSL) is a hyperparameter of the model that limits the maximum size of a single input. According to our experience, if the MaxSL is too short, the input will be truncated, which can negatively impact the performance of representation learning. Conversely, if the MaxSL is too long, it will increase the cost of both model training and usage. This study evaluated four different MaxSLs: Seq128, Seq256, Seq512, and Seq1024. 

\begin{figure}[!t]
    \centering
    \includegraphics[width=4in]{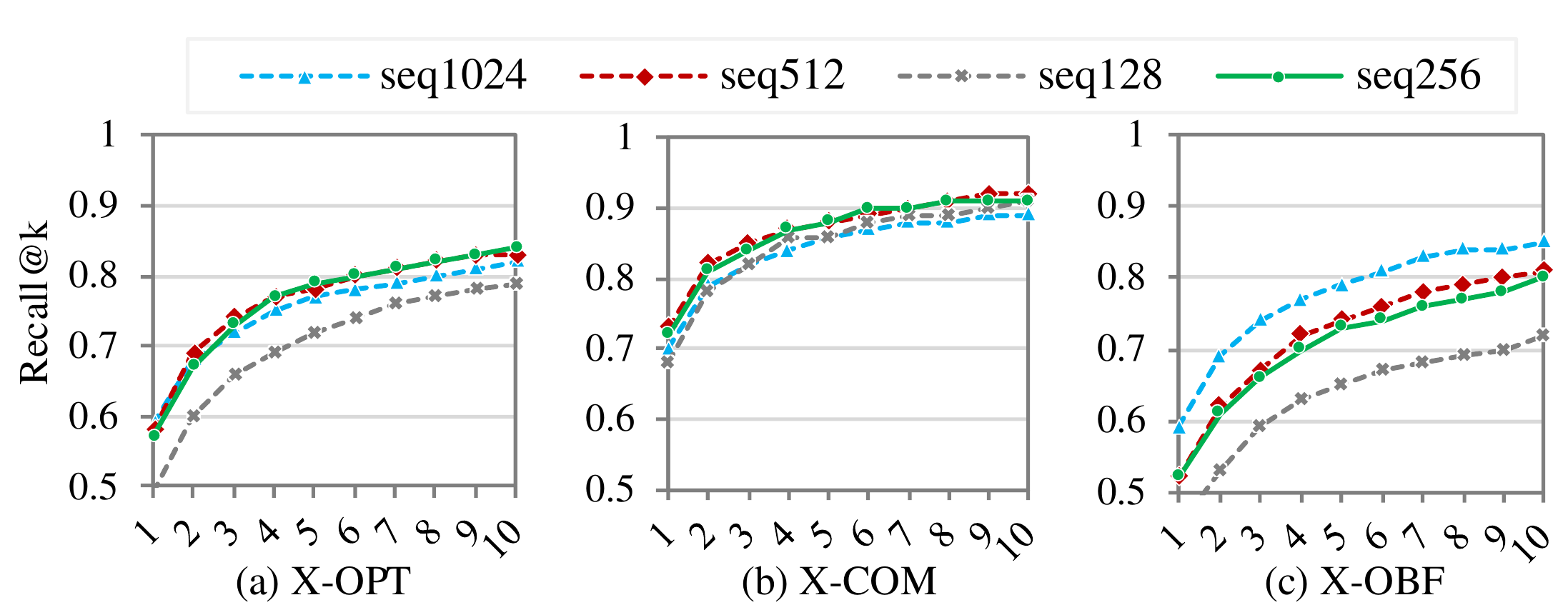}
    \caption{Performance of different max sequence lengths.}
    \label{fig:perf seqlen}
\end{figure}

\begin{figure}[!t]
    \centering
    \includegraphics[width=3.5in]{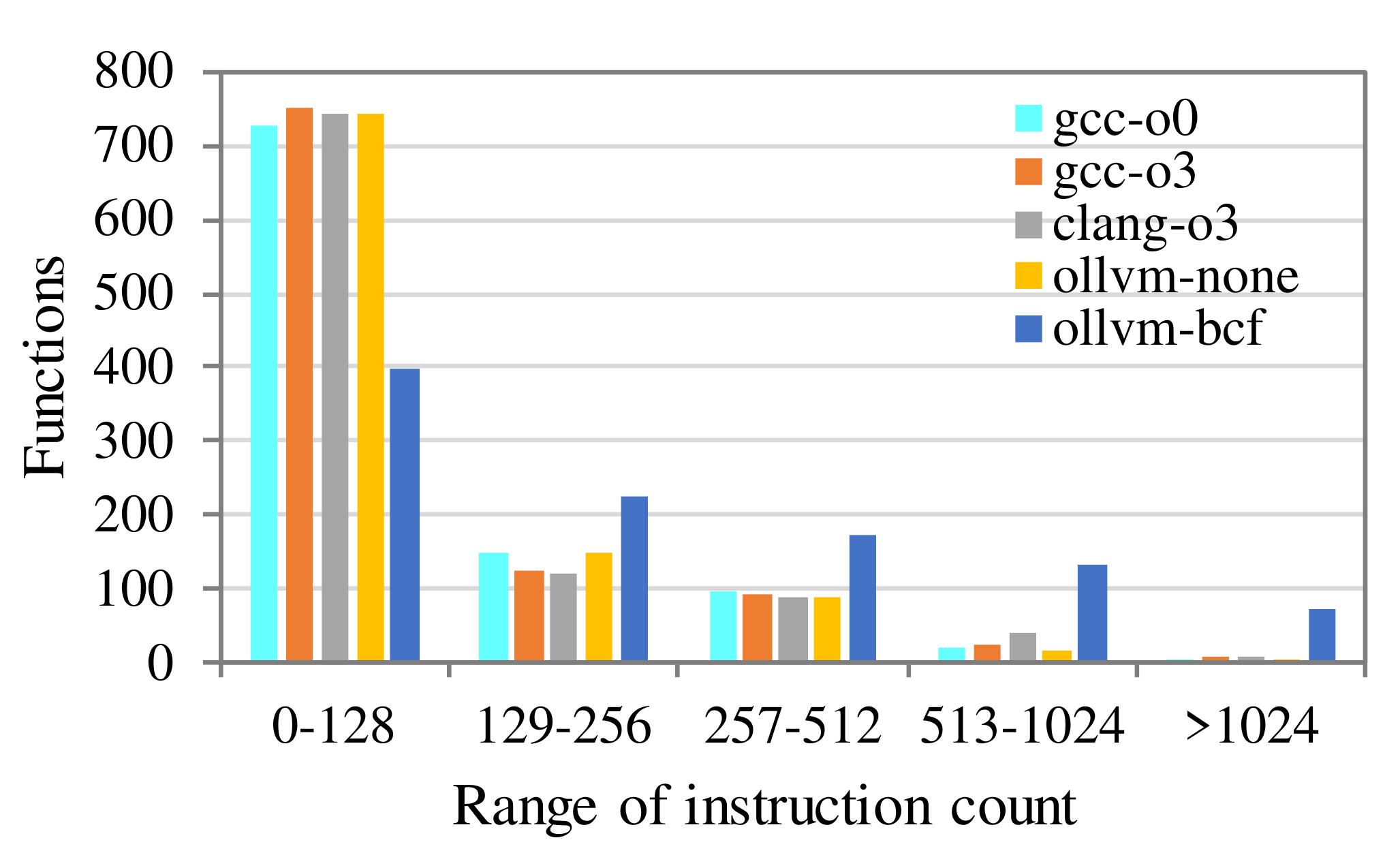}
    \caption{Functions across different ranges of instruction counts.}
    \label{fig:perf seqlen instr count}
\end{figure}

The results (Figure \ref{fig:perf seqlen}) show that Seq128 performs worse than the others in all tasks, and Seq1024 performs significantly better than the others only in the X-OBF task. The possible reason for this phenomenon is that the input length did not exceed the MaxSL. To verify it, we counted the number of instructions in the functions in DS-OBF used in this study. As shown in Figure \ref{fig:perf seqlen instr count}, all functions were placed into five buckets based on their instruction counts: [0-128], [129-256], [257-512], [513-1024], and [\textgreater 1024]. We find that about 35\% of the functions contain more than 128 instructions, which makes Seq-128 suffer a serious truncation problem, resulting in decreased performance. However, only about 12\% of the functions contain more than 256 instructions, making Seq256, Seq512, and Seq1024 perform similarly in X-OPT and X-COM tasks. For the X-OBF task, the “bcf” obfuscation algorithm inserts bogus control flow into the functions. The increased number of instructions enables Seq1024 to leverage its advantages better.

In summary, longer MaxSL has an advantage in handling larger functions. According to the statistics, most functions contain fewer than 256 instructions. Therefore, this paper uses Seq256 as the default MaxSL, which can achieve acceptable performance at a lower cost.

\begin{figure*}[!t]
    \centering
    \includegraphics[width=5.4in]{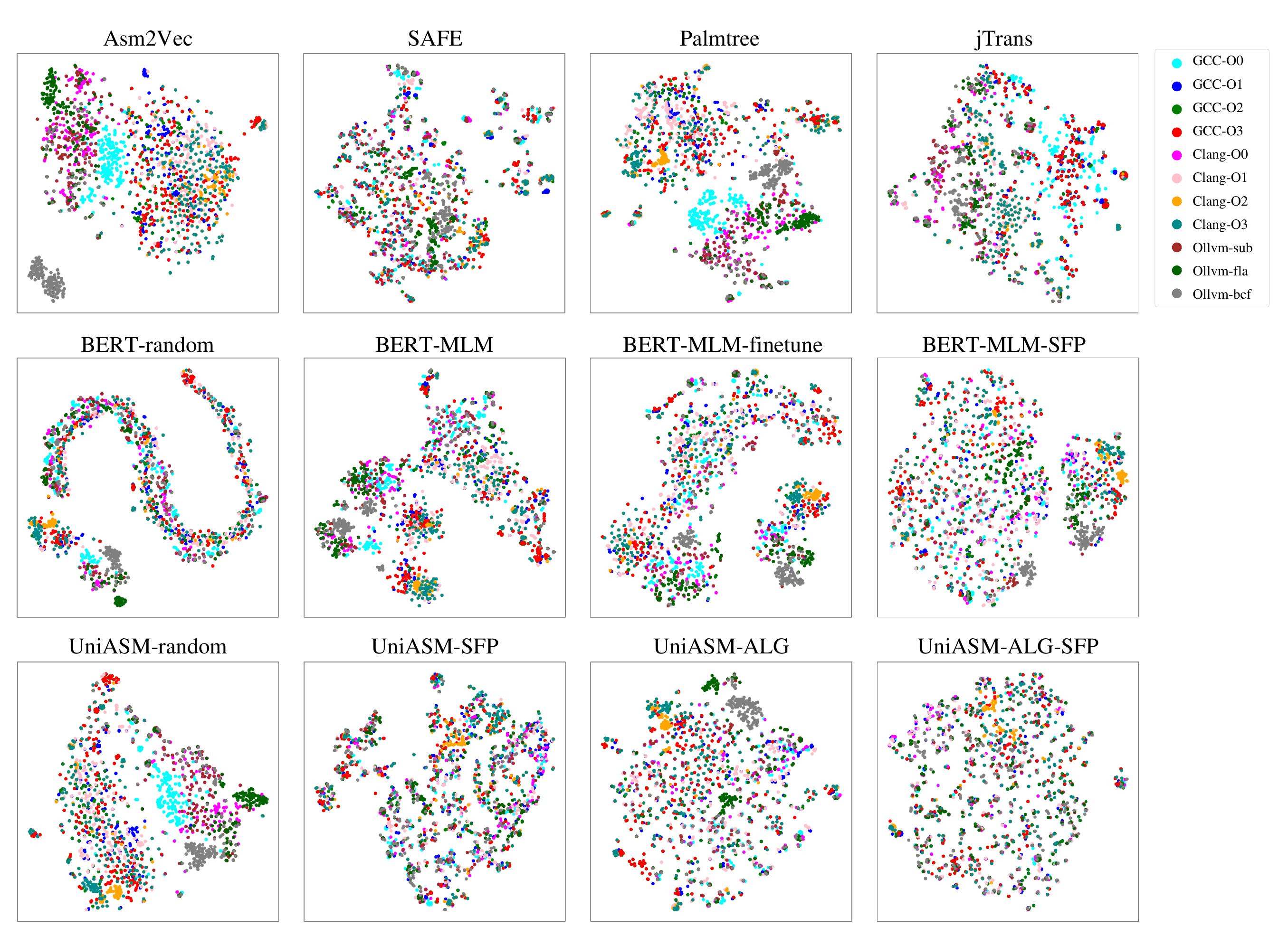}
    \caption{Embedding space of different models.}
    \label{fig:perf space}
\end{figure*}

\subsubsection{Embedding Space Analysis}
\label{sec:abla space}

The function embeddings were all generated from DS-OBF. We compared BERT and UniASM with different training tasks to show the impact of the tasks. “-random” means that the model uses the initial random parameters. In addition, four best-performing baseline models (Asm2Vec, SAFE, Palmtree, and jTrans) were selected for comparison. We leveraged t-SNE \cite{Maaten2008VisualizingDU} to visualize the high-dimensional vectors. Each color indicates one compilation environment.

When calculating similarity, we want similar embeddings to be as close as possible and different embeddings as far as possible. As all functions of a compilation environment (the points in the same color) are considered different, a good embedding space should make the points uniformly distributed.

As shown in Figure \ref{fig:perf space}, the embeddings of UniASM are more uniformly distributed than BERT, which explains why UniASM performs better than BERT. The embedding spaces have significant differences when UniASM is trained with different tasks. Applying SFP alone does not distinguish the embeddings very well. ALG does better, but there are still some local clusters. The joint task ALG+SFP makes most of the embeddings uniformly distributed. It is worth mentioning that BERT-MLM-finetune is the BERT model fine-tuned by a similarity classification task. According to our testing, although it can handle classification tasks well, the generated function embeddings perform poorly in BCSD tasks. Overall, the results of this study are very consistent with the previous evaluation results.

\begin{table*}[!t]
\renewcommand{\arraystretch}{1.2}
\caption{Performance on Vulnerability Searching}
\label{tab:perf vuln}
\centering
\small
\begin{tabular}{cccccccc}
\hline
\multirow{2}{*}{Vulnerability}  & \multirow{2}{*}{Pool} & \multicolumn{5}{c}{Models (Mean Recall@11)} \\ \cline{3-7} 
 &  & Asm2vec & SAFE & PalmTree & jTrans & UniASM\\
\hline
CVE-2013-1944  & 8334 & .16 & .27 & .14 & .27 & \textbf{.77}\\
CVE-2015-8877  & 4296 & .33 & .36 & .22 & .36 & \textbf{.69}\\
CVE-2016-1541  & 15125 & .25 & .22 & .22 & .28 & \textbf{.86}\\
CVE-2016-7163  & 4804 & .17 & .38 & .26 & .24 & \textbf{.45}\\
CVE-2016-8858  & 53454 & .12 & .21 & .16 & .23 & \textbf{.42}\\
CVE-2017-9051  & 23048 & .21 & .21 & .20 & .38 & \textbf{.54}\\
CVE-2017-7866  & 96836 & .21 & .36 & .36 & .28 & \textbf{.52}\\
CVE-2018-8970  & 50762 & .21 & .27 & .21 & .18 & \textbf{.35}\\
\hline
\end{tabular}
\end{table*}

\subsection{Vulnerability Searching}
\label{sec:eval vuln}

Vulnerability searching is one of the main applications in computer security. This evaluation compared UniASM’s performance with four best-performing baseline models (Asm2Vec, SAFE, Palmtree, and jTrans). The evaluation dataset is DS-VUL, detailed in Section \ref{sec:eval dataset}. For each vulnerability query, the source function pool is the 11 vulnerable function variants, and the target function pool is all functions in all variants. The size of the target function pool for each project varies from 4,296 to 96,836. For example, the target function pool of CVE-2013-1944 from the curl-7.29.0 project contains 8334 functions.

As the source pool contains 11 vulnerable functions, we query each function in the target function pool and collect the top-11 results. Recall@11 was used as the evaluation metric, meaning the model retrieves how many vulnerable functions in the top-11 results. We calculated the mean Recall@11 for all 11 queries. As shown in Table \ref{tab:perf vuln}, UniASM outperforms all baseline models, and its score is 29\% to 207\% higher than the leading baseline.

\section{Limitations}
\label{sec:limit}

In this section we discuss some limitations of our model:

\textbf{Cross-Architecture} As the training dataset of UniASM consists of x86\_64 code. Our pre-trained model can only be used for x86\_64 binaries. However, UniASM is not limited to this and can be re-trained with the dataset of other architectures (e.g., ARM, MIPS, etc.).

\textbf{Control flow semantics} UniASM performs linear serialization of functions, so the current model cannot learn the control flow semantics. Although our ablation studies show that the linear one is similar to the random-walk or the longest-walk. Existing work, such as jTrans, shows that control flow information is an important semantic component of functions. A reasonable representation of the control flow should be helpful and deserves further study.

\textbf{Out-of-vocabulary} Our tokenizer treats the whole instruction as a token, which makes the token contain more semantics information. However, a more complex token means a larger dictionary, leading to the OOV problem. In this paper, UniASM tries to mitigate the OOV problem by normalizing the instructions.

\section{Conclusion and Future Work}
\label{sec:conclustion}

In this paper, we propose UniASM, the first attempt to apply an UniLM-based model to BCSD with two fine-designed training tasks. UniASM learns the semantics of assembly code and generates the function embeddings. The generated vectors can be used directly for similarity comparisons without fine-tuning. Experimental results show that UniASM has better performance than the top-performing baselines. In addition, we conduct ablation studies to explore the factors that affect the model’s accuracy in BCSD tasks.

ALG gives the model the ability to generate assembly code. In the future, we plan to apply this ability to more valuable downstream tasks, such as code transformation, automatic coding, etc.

\section*{Acknowledgment}

The authors would like to thank the anonymous reviewers for their insightful comments on our work. Any opinions, findings, and conclusions or recommendations expressed in this paper are those of the authors and do not necessarily reflect the views of the funding agencies.

\section*{Funding Statement}
This paper is supported by the National Key Research and Development Project (2019QY1305).

\section*{Conflicts of Interest}
The authors declare that they have no conflicts of interest to report regarding the present study.

\appendix

 \bibliographystyle{elsarticle-num} 
 \bibliography{uniasm}





% \end{thebibliography}
\end{document}